\documentclass[11pt,a4paper,american,british]{article}
\pdfoutput=1
\usepackage{lmodern}

\usepackage[]{fontenc}
\usepackage[utf8]{inputenc}
\usepackage{color}
\usepackage{booktabs}
\usepackage{units}
\usepackage{amsmath}
\usepackage{amssymb}
\usepackage{esint}

\makeatletter

\pdfpageheight\paperheight
\pdfpagewidth\paperwidth

\providecommand{\tabularnewline}{\\}

\usepackage{jcappub}

\renewcommand{\[}{\begin{equation}}
\renewcommand{\]}{\end{equation}} 
\usepackage{array}

\makeatother

\usepackage{babel}
\begin{document}

\title{Maximal freedom at minimum cost: linear large-scale structure in
general modifications of gravity}

\author[a]{Emilio Bellini}

\author[b]{and Ignacy Sawicki}

\affiliation[a]{Institut für Theoretische Physik, Ruprecht-Karls-Universität Heidelberg\\
Philosophenweg 16, 69120 Heidelberg, Germany}

\affiliation[b]{African Institute for Mathematical Sciences\\
7 Melrose Road, Cape Town 7945, South Africa}

\emailAdd{bellini@thphys.uni-heidelberg.de, ignacy.sawicki@outlook.com}

\abstract{We present a turnkey solution, ready for implementation in numerical
codes, for the study of linear structure formation in general scalar-tensor
models involving a single universally coupled scalar field. We show
that the totality of cosmological information on the gravitational
sector can be compressed --- without any redundancy --- into five
independent and arbitrary functions of time only and one constant.
These describe physical properties of the universe: the observable
background expansion history, fractional matter density today, and
four functions of time describing the properties of the dark energy.
We show that two of those dark-energy property functions control the
existence of anisotropic stress, the other two --- dark-energy clustering,
both of which are can be scale-dependent. All these properties can
in principle be measured, but no information on the underlying theory
of acceleration beyond this can be obtained. We present a translation
between popular models of late-time acceleration (e.g.~perfect fluids,
$f(R)$, kinetic gravity braiding, galileons), as well as the effective
field theory framework, and our formulation. In this way, implementing
this formulation numerically would give a single tool which could
consistently test the majority of models of late-time acceleration
heretofore proposed.}

\maketitle

\section{Introduction\label{sec:Intro}}

Even before the discovery of the acceleration of the expansion of
the universe, dynamical mechanisms for generating such a phase in
the evolution history were being proposed (e.g.~quintessence \cite{Ratra:1987rm,Wetterich:1987fm}).
The actual discovery resulted in a very large volume of proposals
for dynamical models of this late-time acceleration, both based on
classical field theories (usually of a scalar field, but also modified
theories of the gravitational metric, e.g.~$f(R)$ gravity \cite{Carroll:2003wy})
and on phenomenological approaches (e.g.~see reviews~\cite{Clifton:2011jh,Kunz:2012aw}).
It was quickly understood that one could potentially get a handle
on the actual mechanism behind the acceleration by comparing the behaviour
of the background expansion history with that of the large-scale structure
which evolves on this background \cite{Ishak:2005zs}.

It is really this comparison that is key to understanding the nature
of dark energy. A detection of an expansion history incompatible with
a $\Lambda$-cold-dark-matter ($\Lambda$CDM) universe does mean that
an explanation alternative to the cosmological constant is required,
but does not actually say much about the nature of the dynamical mechanism.
Conversely, the $\Lambda$CDM framework describes the whole gravitational
sector using one constant parameter ($\Omega_{\text{\ensuremath{\Lambda}}}$)
measurable from the background expansion. The growth of structure
is then completely fixed (the remaining theoretical uncertainties
on non-linear scales notwithstanding) and a measurement of the properties
of large-scale structure is \emph{necessary }to ascertain that some
other mechanism with background expansion close to $\Lambda$CDM is
not actually responsible. Just measuring the background is not enough.

Structure in the can be studied universe in two approaches. One can
build model-independent null tests from observables which can be used
to constrain properties of the models. This approach is well studied
for the case of the evolution of the background \foreignlanguage{american}{\cite{Huterer:2002hy,Clarkson:2007pz,Zunckel:2008ti}}.
In the case of perturbations, it was demonstrated that one can in
principle reconstruct the gravitational potentials as a function of
time and scale and therefore measure such properties as the anisotropic
stress in a model-independent manner \cite{Amendola:2012ky,Motta:2013cwa}.
For example, Euclid should be able to measure the slip parameter in
such a model-independent manner to within a few percent, depending
on the assumptions \cite{Amendola:2013qna}. 

A much more usual approach is to take a particular dark energy model
and fold in the evolution described by it into a full cosmological
Boltzmann code, such as CAMB \cite{Lewis:1999bs} or CLASS \cite{Lesgourgues:2011re,Blas:2011rf}
following this with a Monte Carlo exploration of the parameter space
allowed by the totality of all the available observational data. The
disadvantage of this method is that it is very computationally expensive
since each model must be constrained separately, unless it is nested
in some sort of larger model class. Typically one therefore reduces
the models to a small number of parameters which capture the gross
behaviour and tries to constrain the parameters' allowed values. For
example, it is well understood that, on the level of the background,
the effective equation of state of the dark energy, which is in principle
a free function of time, fully describes the evolution of the metric.
It is frequently parameterised using the first two terms of a Taylor
expansion \cite{Chevallier:2000qy,Linder:2002et}. 

The description of perturbations is more complicated. Scalar perturbations
dominate at linear scales and, in principle, all one needs to do is
to relate the two scalar gravitational potentials to the matter distribution
through some sort of parameterisation involving two functions of both
scale and time \cite{Caldwell:2005ai,Linder:2005in,Amendola:2007rr}.
The remaining question is how to do it in a way that models properties
of the underlying dark-energy theory, rather than being an arbitrary
parameterisation which could capture a deviation away from the concordance
model but not necessarily the physics. To be specific, general relativity
has a generally covariant structure with no scalar degrees of freedom
of its own, which leads to a particular set of relations between the
gravitational potentials, their derivatives, and the density and velocity
perturbations of the matter. The idea is to find the appropriate ways
of generalising such relationships in more complicated theories.

For example, one can appropriately take large-scale effects into consideration
\cite{Bertschinger:2006aw}, but this is clearly not enough to fully
specify the theory. This can be matched to small-scale behaviour using
a phenomenological interpolation \cite{Hu:2007pj,Lombriser:2013aj}.
Alternatively, requiring gauge invariance of the equations \cite{Baker:2011jy,Baker:2012zs}
restricts the possible structures, but still leaves a very large number
of free functions to be determined. One can then attempt to describe
the dark energy through its field content and its effective fluid
--- or some sort of hopefully physical --- properties \cite{Sawicki:2012re,Battye:2012eu,Battye:2013aaa,Battye:2013ida}.
Instead, one can assume that the dark-energy model preserves locality
and is approximately quasi-static, which results in a particular behaviour
of the perturbations with scale \cite{Bertschinger:2008zb,Amendola:2012ky,Silvestri:2013ne,Hojjati:2013xqa}. 

An alternative approach is to consider an effective field theory (EFT)
for perturbations on top of a Friedmann-Robertson-Walker (FRW) cosmological
background. This method was introduced in \cite{Creminelli:2006xe}
and has been developed in the context of general dark-energy cosmologies
in \cite{Gubitosi:2012hu,Bloomfield:2012ff,Gleyzes:2013ooa,Bloomfield:2013efa,Piazza:2013pua,Gergely:2014rna}
(see also the review \cite{Tsujikawa:2014mba}). The idea is that
one writes down an action for perturbations containing all possible
operators allowed by the symmetries of the FRW background and properties
required of the underlying dark energy. These operators have arbitrary
coefficients that are purely functions of time and represent the maximal
freedom that is permitted in the equations of motion for perturbations.
Not all these operators contribute in a distinct manner to the perturbations
\cite{Piazza:2013pua} but they do represent the maximal freedom that
one has in deforming Einstein equations according to whichever DE
properties are required. This approach to perturbations has been implemented
numerically in ref.~\cite{Hu:2013twa}.

Our approach in this paper is similar in spirit to the EFT approach,
although we restrict our attention to the Horndeski class of models
\cite{Horndeski:1974wa}. The Horndeski models are the most general
set of theories of a single extra scalar degree of freedom or, equivalently,
which have equations of motion with at most second derivatives on
any background \cite{Deffayet:2011gz}. The majority of universally
coupled dark energy and modified gravity models belong to this class.
We show that the evolution of linear perturbations can be completely
described by specifying an arbitrary background evolution history,
the constant $\Omega_{\text{m}0}$ --- fractional matter density today
--- and four additional independent functions of time only which fully
specify the effects of the dark-energy model. We provide the full
set of equations for direct implementation in standard cosmological
codes. Since most popular models of dynamical dark energy fall into
the scope of this paper, they can be easily mapped onto our formulation.
We present this mapping, demonstrating how a code modified for our
formulation can immediately be used to model perturbations in these
less general classes of dark-energy theories, allowing for a single
code to be used to constrain many classes of models.

Our choice of these four functions is equivalent to a particular combination
of the EFT operators of refs~\cite{Gubitosi:2012hu,Bloomfield:2012ff},
but one which has the advantage of mapping directly onto physical
effects. In addition, contrary to the EFT descriptions
where some of the operators enter both the background and the perturbation
properties, our formulation completely separates the two. We also
explicitly demonstrate that the four two of these parameters control
the existence of anisotropic stress while the other two control the
clustering of the dark energy

A non-detection of anisotropic stress or any deviation from the standard
growth rate would constrain the parameters to lie close to their $\Lambda$CDM
values of zero. Combined with an expansion history consistent with
$\Lambda$CDM, such a result would imply that any dynamical dark-energy
mechanism is responsible for no more than a small fraction of the
acceleration, with the remainder being driven by the standard vacuum
energy.

We begin in section \ref{sec:Scope} by describing the scope of our
approach. In section~\ref{sec:DynDesc}, we define our formulation
and present the full background and perturbation equations in Newtonian
gauge in terms of our parameter functions. We discuss the requirements
posed by stability considerations, connect our formulation to previous
approaches and provide a mapping between popular models of dark energy
onto our formulation. We discuss the effect of the dark energy on
the physics of structure formation in section~\ref{sec:Physics},
discussing the validity of the quasi-static limit, constraints from
Solar-System tests and the measurability of the dark energy parameters.
We discuss our findings and conclude in section~\ref{sec:concl}.
In the Appendix, we present the relevant equation in synchronous gauge,
allowing for immediate inclusion in popular Boltzmann codes.

\section{Scope: General Scalar-Tensor Theories\label{sec:Scope}}

In this paper, we discuss the properties of linear perturbations of
scalar-tensor theories belonging to the Horndeski class of actions
\cite{Horndeski:1974wa,Deffayet:2011gz} when evolving on a cosmological
background. This action is the most general action for a single classical
scalar field in the presence of gravity which does not result in any
derivatives higher than second order in the equations of motion on
any metric. We assume that the weak equivalence principle holds and
therefore that all matter species external to the scalar-tensor system
are coupled minimally and universally. 

The combined action for gravity and the scalar is defined by

\begin{equation}
S=\int\mathrm{d}^{4}x\,\sqrt{-g}\left[\sum_{i=2}^{5}{\cal L}_{i}\,+\mathcal{L}_{\text{m}}[g_{\mu\nu}]\right]\,,\label{eq:lagrangian}
\end{equation}
where $g_{\mu\nu}$ is the metric to which the matter sector, described
by $\mathcal{L}_{\text{m}}$, is coupled.%
\footnote{We use the $(-+++)$ signature convention.%
} For the purpose of this paper, the matter sector should be thought
of as describing all of dark matter, baryons, radiation and neutrinos;
we will not differentiate between them here. The Lagrangians ${\cal L}_{i}$
are usually written as

\begin{eqnarray*}
{\cal L}_{2} & = & K(\phi,\, X)\,,\\
{\cal L}_{3} & = & -G_{3}(\phi,\, X)\Box\phi\,,\\
{\cal L}_{4} & = & G_{4}(\phi,\, X)R+G_{4X}(\phi,\, X)\left[\left(\Box\phi\right)^{2}-\phi_{;\mu\nu}\phi^{;\mu\nu}\right]\,,\\
{\cal L}_{5} & = & G_{5}(\phi,\, X)G_{\mu\nu}\phi^{;\mu\nu}-\frac{1}{6}G_{5X}(\phi,\, X)\left[\left(\Box\phi\right)^{3}+2{\phi_{;\mu}}^{\nu}{\phi_{;\nu}}^{\alpha}{\phi_{;\alpha}}^{\mu}-3\phi_{;\mu\nu}\phi^{;\mu\nu}\Box\phi\right]\,.
\end{eqnarray*}
The couplings to gravity are completely fixed by the Lagrangians $\mathcal{L}_{4-5}$.
The four functions $K(\phi,X)$ and $G_{i}(\phi,X)$ are arbitrary
functions of the scalar field $\phi$ and its canonical kinetic term
$X=-\phi^{;\mu}\phi_{;\mu}/2$. The subscript $X$ represents a derivative
w.r.t.\ $X$ while $\phi$ is a derivative w.r.t.\ the field $\phi$.
This class of models therefore comprises four functional degrees of
freedom which will combine into the four degrees of freedom in our
parameterisation. 

This general class of actions includes essentially all universally
coupled models of dark energy with one scalar degree of freedom: quintessence
\cite{Wetterich:1987fm,Ratra:1987rm}, Brans-Dicke models \cite{Brans:1961sx},
k-\emph{essence} \cite{ArmendarizPicon:1999rj,ArmendarizPicon:2000ah},
kinetic gravity braiding \cite{Deffayet:2010qz,Kobayashi:2010cm,Pujolas:2011he},
galileons \cite{Nicolis:2008in,Nicolis:2009qm}. Archetypal modified-gravity
models such as $f(R)$ \cite{Carroll:2003wy} and $f(G)$ \cite{Carroll:2004de}
gravity are within our purview. Models such as Dvali-Gabadadze-Porrati
(DGP) gravity \cite{Dvali:2000hr} or ghost-free massive gravity \cite{deRham:2010ik,deRham:2010kj,Hassan:2011tf}
lie outside of our purview. However, their appropriately covariantised
decoupling limits, Galileon Cosmology \cite{Chow:2009fm} and proxy
massive gravity \cite{deRham:2011by,Heisenberg:2014kea}, which should
correctly capture the behaviour of the scalar degree of freedom, do
belong to the Horndeski class and therefore are described by our formulation.

As we explain later, our approach is also closely related to the effective
field theory method of describing linear perturbations on the Friedman
cosmological metric. It was proven in ref.~\cite{Gleyzes:2013ooa}
that the maximal theory for linear perturbation described by equations
with no more than second derivatives is equivalent to that given by
the Horndeski action. Although ref. \cite{Gleyzes:2013ooa} notes
that there is an extension allowed where the Einstein equations contain
third derivatives, but the equation for the propagating scalar degree
of freedom does not. Surprisingly, this structure seems to be repeated
on non-linear level \cite{Gleyzes:2014dya}.%
\footnote{See also the possibly related ref.~\cite{Zumalacarregui:2013pma}.%
}

Some models of dark energy break the equivalence principle between
the baryonic and dark sectors (e.g.~coupled quintessence \cite{Amendola:1999er}
or non-universal disformal couplings \cite{Koivisto:2012za,Zumalacarregui:2012us})
allowing, for example, for a much simpler evasion of Solar-System
constraints. Our method could be extended to these models quite simply
with the introduction of an additional parameter for the coupling
to dark matter and the relative DM/baryon density. We must note that
our framework does not in general cover such classes of models as:
Lorentz-invariance-violating models (e.g.~Hořava-Lifschitz models
\cite{Horava:2009uw,Blas:2009qj}) or non-conservative fluids arising
from high-temperature self interactions of a scalar (e.g. Dark Goo
\cite{Gagnon:2011id}).

\section{Minimal Description of Dynamics\label{sec:DynDesc}}

We assume that the universe is well described by small scalar perturbations
on top of a FRW metric. We assume spatial flatness and use the Newtonian
gauge with the notation of ref.~\cite{Ma:1995ey}
throughout the paper%
\footnote{Note that in this paper we used $\Psi$ and $\Phi$ instead of $\psi$
and $\phi$, being these potentials the correspondent gauge invariant
variables in the Newtonian gauge.%
}, i.e.~the line element takes the form
\begin{equation}
\mathrm{d}s^{2}=-(1+2\Psi)\mathrm{d}t^{2}+a^{2}(t)(1-2\Phi)\mathrm{d}\boldsymbol{x}^{2}\,.\label{eq:Metric}
\end{equation}
We have chosen this gauge since the metric potentials $\Phi$ and
$\Psi$ are in fact cosmological observables \cite{Motta:2013cwa}.
We present the main results in synchronous gauge in the Appendix \ref{sec:SynchGauge}.

The purpose of the presentation that follows in the rest of this section
is dual. Firstly, it provides sufficient information to fully solve
for the evolution of linear cosmological perturbations in any particular
dark-energy model in the Horndeski class. Given an action for a model,
all one is required to do is to compute the form of four functions
$\alpha_{i}$ defined in section~\ref{sub:omegas}
and solve the resulting equations of motion for the background and
perturbations. On this level, the method gives a unified, ``write
once use many'' approach for evolving linear large-scale structure
in dark energy models. To obtain constraints on parameters of the
Lagrangian, the algorithm would need to scan through potential initial
conditions on the scalar field value, solving for the background evolution
and then determining the values of the $\alpha_{i}$ as a function
of time for that model. This is what is being done today for example
for covariant galileon models \cite{Barreira:2012kk}; our formulation
essentially provides a book-keeping device when seen from this perspective.

However, a potentially more fruitful approach would be one close to
that taken by EFT methods: observations to date have determined that
the background expansion history appears close to $\Lambda$CDM. Since
our description identifies the maximal number of functions independent
of background expansion, the equations we provide can be used to constrain
the forms of these property functions $\alpha_{i}$ compatible with
the observed quasi-$\Lambda$CDM background, without making any reference
to a particular model or initial conditions for the scalar's background
value. In fact, the values of the functions $\alpha_{i}$ are the
maximum unambiguous information that can ever be extracted about such
dark-energy models from the evolution of linear cosmological perturbations.

We now describe the physical meaning and origin of
the four property functions $\alpha_{i}$, before turning to the background
equations and stability tests and finally the full perturbations equations
involving all of these property functions.

\subsection{Non-Redundant Set of Linear-Perturbation Properties\label{sub:omegas}}

It is well known that the evolution history of the cosmological background
does not specify the dark-energy model (for example quintessence and
$f(R)$ gravity have different predictions for the growth of structure
given the same expansion history). The main result of this paper is
that we have formulated the possible linear perturbation theory for
dark-energy models belonging to class \eqref{eq:lagrangian} in terms
of \emph{four} functions of time, $\alpha_{i}(t)$. Augmenting them
with the background evolution history $H(t)$ and the matter density
today $\tilde{\rho}_{\text{m0}}$ \emph{fully }determines the evolution
of large-scale structure. It is important to stress that for a general
model all of these are completely independent. Equally importantly:
this is the \emph{minimal }set of functional parameters which is capable
of describing all models within our scope. 

When considering a particular model, with appropriate initial conditions
for the background variables, the four functions $\alpha_{i}$ are
determined by the Lagrangian and the value of the scalar field through
the definitions (\ref{eq:omega1-1}-\ref{eq:omega4}). On one hand
they can be considered a helpful compression of the perturbations
equations. However, this is not the approach we advocate here. Given
a fixed background and $\tilde{\rho}_{\text{m}0}$, any two trajectories
in any two seemingly different models which have the same $\alpha_{i}$
exhibit \emph{absolutely }no difference in their behaviour. This implies
that no measurements of linear structure can \emph{ever }say anything
about the particular theory of dark energy beyond the measurement
of $\rho_{\text{m}0}$ and $\alpha_{i}(t)$. This is the maximum extent
to which dark-energy Horndeski Lagrangians can ever be reconstructed
using linear-theory observations.

The above is strictly true only when the scalar rolls monotonically,
i.e.~$\dot{\phi}\neq0$. If there are oscillations of the \emph{background}
values (e.g.~see ref.~\cite{Amin:2011hu}), then the equations need
to be expanded and they do depend also on the value of $\dot{\phi}$.
In such a case, the full model must be supplied and solved for.

Since the four DE property functions $\alpha_{i}$ are arbitrary and
independent of the background, the matter sector and of each other,
we argue that they should be thought of as being essentially the relevant
independent physical properties of the dark energy and should be the
ones targeted for measurement. We describe the physical
meaning of these function here, while their analytic definition in
terms of the Horndeski functions $K$ and $G_{i}$ can be found in
Appendix~\ref{sec:Definitions-of-Evolution},
\begin{itemize}
\item $\alpha_{\textrm{K}}$, \emph{kineticity. }Kinetic energy of scalar
perturbations arising directly from the action. It is the only contribution
of perfect-fluid models but is not present at all in archetypal ``modified
gravity'' models such as $f(R)$ and $f(G)$. Large
values act to suppress the sound speed of scalar perturbations. Contribution
from all of $K$, $G_{3}$, $G_{4}$, $G_{5}$.
\item $\alpha_{\textrm{B}}$, \emph{braiding.} Signifies braiding, or mixing
of the kinetic terms of the scalar and metric. Contributes to kinetic
energy of scalar perturbations indirectly, by backreacting through
gravity. Second time derivatives of metric and scalar field to appear
in equations of motion for both the scalar and the metric. Causes
dark energy to cluster. Contributions from $G_{3}$, $G_{4}$, $G_{5}$.
\item $\alpha_{\textrm{M}}$, \emph{Planck-mass run rate.} Rate of evolution
of the effective Planck mass. A constant redefinition of the Planck
mass does not affect physics. Its time evolution in the Jordan frame
of the observer creates anisotropic stress. Contributions from $G_{4}$
and $G_{5}$.
\item $\alpha_{\textrm{T}}$, \emph{tensor speed excess.} Deviation of the
speed of gravitational waves from that of light. This violation of
Lorentz-invariance for tensors also changes the response of the Newtonian
potential $\Psi$ to matter sources even in the presence of no scalar
perturbations, leading to anisotropic stress. Contributions from $G_{4}$
and $G_{5}$.
\end{itemize}
The existence of these four contributions can be roughly allocated
to the four possible types of perturbation structures in the dark-energy
EMT at linear order. Following ref.~\cite{Sawicki:2012re}, we analyse
the perturbations on ``comoving'' $\phi=\text{const}$ (equal clock)
hypersurfaces. 
\begin{itemize}
\item The $\alpha_{\textrm{K}}$ contributions are essentially those of
linear perturbations of the standard perfect-fluid EMT. 
\item Braiding $\alpha_{\text{B}}$ represents a space-like energy flux
vector $q^{i}$ in this frame (a $T^{0i}$ contribution); essentially
this is a generalisation of the imperfect contributions described
for kinetic gravity braiding models in \cite{Pujolas:2011he}. 
\item The change in the speed of tensors $\alpha_{\textrm{T}}$ arises as
a contribution in the DE EMT proportional to the curvature of the
spatial comoving hypersurfaces, $^{(3)}R_{\mu\nu}$, which contains
the graviton gradient terms. The effect of this is to introduce an
offset between the two metric potentials, i.e.\ anisotropic stress.
\item Finally, $\alpha_{\textrm{M}}$ is a result of contributions in the
EMT proportional to the Einstein tensor. On $\phi=\text{const}$ hypersurfaces
at linear order it has no impact. However, any variation of the Planck
mass represents a conformal rescaling of the metric which splits the
two gravitational potentials introducing anisotropic stress.
\end{itemize}
We have have presented a list of $\alpha_{i}$'s for a selection of
popular dark-energy model classes in table~\ref{tab:omegas}. Typically,
any particular dark-energy model class will only turn on one of the
$\alpha_{i}$'s (perfect fluids), or the values of multiple parameters
will be related (e.g.\ $f(R)$). Thus measuring the appropriate values
of the parameters allows for a determination of the class of models
into which the dark energy falls. Standard parameterisations for particular
models can easily be used by finding the appropriate conversion between
them and the $\alpha_{i}$.

\begin{table}
\begin{centering}
\begin{tabular}{llrrrr}
\toprule 
\textsf{\textbf{\scriptsize{}Model Class}} &  & \textsf{\textbf{\scriptsize{}$\boldsymbol{\alpha_{\textrm{K}}}$}} & {\scriptsize{}$\boldsymbol{\alpha_{\textrm{B}}}$} & \textsf{\textbf{\scriptsize{}$\boldsymbol{\alpha_{\textrm{M}}}$}} & \textsf{\textbf{\scriptsize{}$\boldsymbol{\alpha_{\textrm{T}}}$}}\tabularnewline
\midrule
\midrule 
\textbf{\emph{\scriptsize{}$\boldsymbol{\Lambda}$CDM }} &  & \textbf{\emph{\scriptsize{}0}} & \textbf{\emph{\scriptsize{}0}} & \textbf{\emph{\scriptsize{}0}} & \textbf{\emph{\scriptsize{}0}}\tabularnewline
\midrule 
{\scriptsize{}cuscuton ($w_{X}\neq-1)$} & {\scriptsize{}\cite{Afshordi:2006ad}} & {\scriptsize{}0} & {\scriptsize{}0} & {\scriptsize{}0} & {\scriptsize{}0}\tabularnewline
\midrule 
{\scriptsize{}quintessence} & {\scriptsize{}\cite{Wetterich:1987fm,Ratra:1987rm}} & {\scriptsize{}$(1-\Omega_{\text{m}})(1+w_{X})$} & {\scriptsize{}0} & {\scriptsize{}0} & {\scriptsize{}0}\tabularnewline
\midrule 
{\scriptsize{}k-}\emph{\scriptsize{}essence}{\scriptsize{}/perfect
fluid} & {\scriptsize{}\cite{ArmendarizPicon:1999rj,ArmendarizPicon:2000ah}} & {\scriptsize{}$\frac{(1-\Omega_{\text{m}})(1+w_{X})}{c_{\text{s}}^{2}}$} & {\scriptsize{}0} & {\scriptsize{}0} & {\scriptsize{}0}\tabularnewline
\midrule 
{\scriptsize{}kinetic gravity braiding} & {\scriptsize{}\cite{Deffayet:2010qz,Kobayashi:2010cm,Pujolas:2011he}} & {\scriptsize{}$\nicefrac{m^{2}\left(n_{m}+\kappa_{\phi}\right)}{H^{2}M_{\text{Pl}}^{2}}$} & {\scriptsize{}$\nicefrac{m\kappa}{HM_{\text{Pl}}^{2}}$} & {\scriptsize{}0} & {\scriptsize{}0}\tabularnewline
\midrule 
{\scriptsize{}galileon cosmology} & {\scriptsize{}\cite{Chow:2009fm}} & {\scriptsize{}$-\nicefrac{3}{2}\alpha_{\textrm{M}}^{3}H^{2}r_{\text{c}}^{2}e^{2\phi/M}$} & {\scriptsize{}$\nicefrac{\alpha_{\textrm{K}}}{6}-\alpha_{\textrm{M}}$} & {\scriptsize{}$\nicefrac{-2\dot{\phi}}{HM}$} & {\scriptsize{}0}\tabularnewline
\midrule 
{\scriptsize{}BDK} & {\scriptsize{}\cite{Sawicki:2012re}} & {\scriptsize{}$\nicefrac{\dot{\phi}^{2}K_{,\dot{\phi}\dot{\phi}}e^{-\kappa}}{H^{2}M^{2}}$} & {\scriptsize{}$-\alpha_{\textrm{M}}$} & {\scriptsize{}$\nicefrac{\dot{\kappa}}{H}$} & {\scriptsize{}0}\tabularnewline
\midrule 
{\scriptsize{}metric $f(R)$} & {\scriptsize{}\cite{Carroll:2003wy,Song:2006ej}} & {\scriptsize{}0} & {\scriptsize{}$-\alpha_{\textrm{M}}$} & {\scriptsize{}$\nicefrac{B\dot{H}}{H^{2}}$} & {\scriptsize{}0}\tabularnewline
\midrule 
{\scriptsize{}MSG/Palatini $f(R)$} & {\scriptsize{}\cite{Carroll:2006jn,Vollick:2003aw}} & {\scriptsize{}$-\nicefrac{3}{2}\alpha_{\textrm{M}}^{2}$} & {\scriptsize{}$-\alpha_{\textrm{M}}$} & {\scriptsize{}$\nicefrac{2\dot{\phi}}{H}$} & {\scriptsize{}0}\tabularnewline
\midrule 
{\scriptsize{}$f($Gauss-Bonnet$)$} & {\scriptsize{}\cite{Carroll:2004de,DeFelice:2009rw,Kobayashi:2011nu}} & {\scriptsize{}0} & {\scriptsize{}$\frac{-2H\dot{\xi}}{M^{2}+H\dot{\xi}}$} & {\scriptsize{}$\frac{\dot{H}\dot{\xi}+H\ddot{\xi}}{H\left(M^{2}+H\dot{\xi}\right)}$} & {\scriptsize{}$\frac{\ddot{\xi}-H\dot{\xi}}{M^{2}+H\dot{\xi}}$}\tabularnewline
\bottomrule
\end{tabular}
\par\end{centering}

\protect\caption{Parameter functions $\alpha_{i}$ for various classes of models determined
from their actions using definitions~(\ref{eq:planckmass}-\ref{eq:omega4}).
For example in quintessence models, background evolution determines
absolutely everything about linear perturbations, while in k-\emph{essence}
we have an additional degree of freedom, the sound speed. Perfect-fluid
models are described by just the kineticity\textbf{ }$\alpha_{\textrm{K}}$.
On the other hand, the archetypal ``modified gravity'' models such
as $f(R)$ and $f(G)$ have $\alpha_{\textrm{K}}=0$, but turn on
other directions in the space of perturbations. Most simple models
of dark energy are described by a single or possibly two functions
forming a subspace in the full potential theory space spanned by the
$\alpha_{i}$. \label{tab:omegas} }
\end{table}

In principle, it might be possible to directly measure the parameters
$\alpha_{i}$ as a function of redshift given good enough data (see
the null tests of ref.~\cite{Motta:2013cwa} for a method), but most
likely they will have to be inferred from the integrated total of
cosmological structure-formation data. This means that a parameterisation
is necessary, especially if this method is to be implemented in Boltzmann
codes. For a general parameterisation we propose that 
\[
\alpha_{i}=(1-\tilde{\Omega}_{\text{m}})\hat{\alpha}_{i}\,,\qquad\hat{\alpha}_{i}=\text{const}
\]
be used. This sort of parameterisation reflects the fact the values
of $\alpha_{i}$ are driven by the same functions of the scalar field
as the energy density and their derivatives. Thus naively one would
expect them to be of similar size, i.e.~$\hat{\alpha}_{i}\sim1$.
If any of the $\hat{\alpha}_{i}\gg1$ (for example, in a perfect fluid
model with $c_{\text{s}}^{2}\ll1+w_{X}$), then this is a sign that
the energy density and pressure both have some kind of cancellation
which disappears when additional derivatives are taken to form the
$\alpha_{i}$. Thus this kind of situation, which is in principle
perfectly well allowed, can be considered evidence of tuning in the
structure of the Lagrangian. Finally, the situation where all the
$\hat{\alpha}_{i}\ll1$ implies that the acceleration is mainly driven
by a cosmological constant, with a marginal contribution from the
dynamical dark-energy mechanism.

\subsection{Cosmological Background\label{sub:Background}}

The Friedmann equations can be obtained in the usual manner. The ambiguity
is the fact that in Horndeski theories the Planck mass can be a function
of time. In particular, the role of the Planck mass is played by
\begin{equation}
M_{*}^{2}(\phi,X,H)\equiv2\left(G_{4}-2XG_{4X}+XG_{5\phi}-\dot{\phi}HXG_{5X}\right)\,,\label{eq:MPl}
\end{equation}
which depends on time implicitly through the solution realised by
the universe in $\phi$, $X$ and $H$. If the Planck mass $M_{*}^{2}$
is constant on the solution realised by the Universe then it can have
no effect on observables, since masses can always be redefined to
reabsorb it. The only dependence of physics can be on the (dimensionless)
rate of evolution of the Planck mass which is one of the property
functions described in section \ref{sub:omegas},
\begin{equation}
\alpha_{\textrm{M}}\equiv H^{-1}\frac{\mathrm{d}\ln M_{*}^{2}}{\mathrm{d}t}\,.\label{eq:omega1}
\end{equation}
The Friedmann equations are

\begin{align}
 & 3H^{2}=\tilde{\rho}_{\text{m}}+\tilde{\mathcal{E}}\label{eq:Friedmann}\\
 & 2\dot{H}+3H^{2}=-\tilde{p}_{\text{m}}-\tilde{\mathcal{P}}\nonumber 
\end{align}
where we have kept the left-hand side standard. $\tilde{\rho}_{\text{m}}$
and $\tilde{p}_{\text{m}}$ are the background energy density and
pressure of all the matter components together, while $\tilde{\mathcal{E}}$
and $\tilde{\mathcal{P}}$ are the background energy density and pressure
of the dark energy.%
\footnote{Bear in mind that we have removed some terms that appear in the EMT
by collecting them to form the Planck mass $M_{*}^{2}$, eq. \eqref{eq:MPl}.%
} We also sometimes use the matter equation of state parameter and
the fractional matter density,
\[
w_{\text{m}}\equiv\frac{\tilde{p}_{\text{m}}}{\tilde{\rho}_{\text{m}}}\,,\qquad\tilde{\Omega}_{\text{m}}\equiv\frac{\tilde{\rho}_{\text{m}}}{3H^{2}}\,.
\]
We are using the tilde to indicate that all the quantities are divided
by $M_{*}^{2}$, e.g.
\begin{equation}
\tilde{\rho}_{\text{m}}\equiv\frac{\rho_{\text{m}}}{M_{*}^{2}}\,.\label{eq:rhotilde}
\end{equation}
The full expressions for $\tilde{\mathcal{E}}$ and $\tilde{\mathcal{P}}$
are given in the Appendix \ref{sec:Definitions-of-Evolution}, eqs.~\eqref{eq:density}
and \eqref{eq:pressure}. Here we will only note that the pressure
term contains a dependence on $\ddot{\phi}$, 
\begin{equation}
\tilde{\mathcal{P}}\equiv\tilde{P}(\phi,X,H)-\alpha_{\textrm{B}}\frac{H\ddot{\phi}}{\dot{\phi}}\,,\label{eq:Pressure}
\end{equation}
where $\tilde{P}$ is the part of the pressure only depending on $\phi$,
$X$ and $H$. The dependence on $\ddot{\phi}$ results from \emph{kinetic
braiding}, the mixing between the scalar and metric kinetic terms
which was extensively described as an essential feature of kinetic
gravity braiding models in Refs~\cite{Deffayet:2010qz,Pujolas:2011he}.
In general Horndeski models, the \emph{braiding }$\alpha_{\textrm{B}}$
receives contributions from all of $G_{3-5}$ and is defined in the
Appendix in eq.~\eqref{eq:omega3}.

As a result of absorbing a potentially time-evolving Planck mass in
the definition of the tilded quantities such as \eqref{eq:rhotilde},
the energy density is not necessarily covariantly conserved,

\begin{align}
 & \dot{\tilde{\rho}}_{\text{m}}+3H\left(\tilde{\rho}_{\text{m}}+\tilde{p}_{\text{m}}\right)=-\alpha_{\textrm{M}}H\tilde{\rho}_{\text{m}}\,,\label{eq:Conservation}\\
 & \dot{\tilde{\mathcal{E}}}+3H\left(\tilde{\mathcal{E}}+\tilde{\mathcal{P}}\right)=\alpha_{\textrm{M}}H\tilde{\rho}_{\text{m}}\,,\nonumber 
\end{align}
but rather is exchanged between the scalar and matter subsystems whenever
the value of $M_{*}^{2}$ is changing. This presents a slight complication:
one must either keep track of the value of $M_{*}^{2}$ but is then
able to use the standard conservation laws for the untilded quantities.
Alternatively, the densities of \emph{all }matter species need to
appropriately integrated whenever $\alpha_{\textrm{M}}\neq0$ using
Eq.\ (\ref{eq:Conservation}). 

It is important to stress that, despite appearing
in background equations, both $\alpha_{\textrm{B}}$ and $\alpha_{\textrm{M}}$
are only possible to determine through their influence on structure
formation. Any expansion history can be realised for any choice of
these two functions. However, the linear perturbation theory changes
when these functions change and this physics cannot be replicated
by changing the other parameters. They are thus linear-perturbation
properties and their appearance in this section is stressed only to
guide those working to constrain models with fully defined actions,
rather than in the effective framework advocated here.

Given the non-conservation of the matter components whenever the Planck
mass runs, there is an ambiguity as to the definition of the equation
of state and its relation to evolution of energy density for the dark
energy. We will choose to make an operational definition based on
the Friedmann equations \eqref{eq:Friedmann} and therefore on the
dark-energy background configuration,
\begin{equation}
1+w_{X}\equiv\frac{\tilde{\mathcal{E}}+\tilde{\mathcal{P}}}{\tilde{\mathcal{E}}}=-\frac{2\dot{H}+\tilde{\rho}_{\text{m}}+\tilde{p}_{\text{m}}}{3H^{2}+\tilde{\rho}_{\text{m}}}\,.\label{eq:wx-def}
\end{equation}
This choice does not in general fully determine the evolution of dark-energy
energy density, but is a useful book-keeping device in the perturbation
equations and stability analysis. We should note that the observation
of the background expansion history \emph{only} determines $H(z)/H_{0}$.
The value of $\tilde{\rho}_{\text{m}0}$, the density of the matter
sector today, is a free parameter that can only be determined through
measurement of large-scale structure \cite{Amendola:2012ky,Motta:2013cwa}.
Depending on different choices of $\tilde{\rho}_{\text{m}0}$, the
history of $w_{X}$ will also differ given a background expansion
history fixed by observations.

Finally, the scalar field value evolves according to the equation
of motion which can be written succinctly as,

\begin{eqnarray}
\dot{n}+3Hn & = & \mathcal{P}_{\phi}\,,\label{eq:BackgEoM}
\end{eqnarray}
where $n$, defined in the Appendix in eq.~\eqref{eq:shiftcharge},
is a shift charge density which is covariantly conserved whenever
the action exhibits a Noether symmetry with respect to constant shifts,
$\phi\rightarrow\phi+\text{const}$. $\mathcal{P}_{\phi}$ is a \emph{partial
}derivative of the pressure with respect to $\phi$, defined by eq.~\eqref{eq:non-shift}.
Whenever the shift symmetry is present, the models feature attractors
where the charge density $n=0$. These provide natural scaling solutions
and post-inflationary conditions for the models \cite{DeFelice:2010pv,Deffayet:2010qz}.

\subsection{Background Stability\label{sub:backStab}}

One of the complications of the non-canonical Horndeski models is
that backgrounds can become unstable to perturbations. The meaning
of these instabilities is that the background solution found is no
longer appropriate. This is a particular concern when no full action
is given, but a more phenomenological approach we advocate here is
taken. Thus one must ensure that instabilities are not present and
discard a particular range of parameters if they are found.

Generally, the background can suffer either from ghost or gradient
instabilities, or both. Gradient instabilities occur when the background
evolves to a region where the speed of sound squared of the perturbations
is negative, leading to an exponential destabilisation of the perturbations
at small scales, with timescales of the order of the cutoff of the
theory. Ghost instabilities occur when the sign of the kinetic term
for of the background perturbations is wrong. Usually, they are discussed
in the context of quantum stability since the vacuum can destabilise
to produce ghost and normal modes, if ghosts are present, dynamical
and coupled (see estimates of the rates of such destabilisation in
\cite{Brax:2014wla}).

The second-order action for perturbations of the Horndeski action
was first derived in ref.~\cite{DeFelice:2011bh} where a 3+1 decomposition
in the unitary gauge to obtain actions for the propagating modes:
the tensors (gravitational waves) $h_{ij}$, the scalar $\zeta$ and
the perfect-fluid matter sound waves, the last of which which we will
not write out here,
\[
S_{2}=\int\mathrm{d}t\mathrm{d}^{3}xa^{3}\left[Q_{\text{S}}\left(\dot{\zeta}^{2}-\frac{c_{\text{s}}^{2}}{a^{2}}(\partial_{i}\zeta)\right)+Q_{\text{T}}\left(\dot{h}_{ij}^{2}-\frac{c_{\text{T}}^{2}}{a^{2}}(\partial_{k}h_{ij})^{2}\right)\right]\,,
\]
with the stability of the background to the scalar modes requiring
\begin{align}
Q_{\text{S}} & =\frac{2M_{*}^{2}D}{(2-\alpha_{\textrm{B}})^{2}}>0\,,\qquad D\equiv\alpha_{\textrm{K}}+\frac{3}{2}\alpha_{\textrm{B}}^{2}\,,\label{eq:scalarstab}\\
c_{\text{s}}^{2} & =-\frac{\left(2-\alpha_{\textrm{B}}\right)\left[\dot{H}-\frac{1}{2}H^{2}\alpha_{\textrm{B}}\left(1+\alpha_{\textrm{T}}\right)-H^{2}\left(\alpha_{\textrm{M}}-\alpha_{\textrm{T}}\right)\right]-H\dot{\alpha}_{\textrm{B}}+\tilde{\rho}_{\textrm{m}}+\tilde{p}_{\textrm{m}}}{H^{2}D}>0\,,\nonumber 
\end{align}
while the stability to the tensor modes requires
\begin{align}
Q_{\text{T}} & =\frac{M_{*}^{2}}{8}>0\,,\label{eq:tensorstab}\\
c_{\text{T}}^{2} & =1+\alpha_{\textrm{T}}>0\,.\nonumber 
\end{align}
Taking all the conditions together also implies that the no-ghost
condition for the scalar perturbations reduces to $D>0$. We should
note here that $c_{\text{s}}$ is the speed of propagation of small
scalar perturbations in the infinite frequency (eikonal) limit, $k\rightarrow\infty$,
and therefore a statement about the causal structure of the model.
It is therefore \emph{not }a function of scale but only of time. It
is \emph{not} in general equivalent to the frequently used function
$C^{2}$ relating the dark-energy pressure and density perturbations,
$\delta\mathcal{P}\equiv C^{2}\delta\mathcal{E}$, which usually depends
on scale, but which is not necessarily related to stability (see e.g.~the
discussion in ref.~\cite{Sawicki:2012re}).

It is the expression for $Q_{T}$ that validates why we chose to call
$M_{*}^{2}$, eq.~\eqref{eq:MPl}, the effective Planck mass: the
unambiguous meaning of Planck mass is as the normalisation of the
graviton kinetic term. The action for the tensor perturbations also
clarifies the meaning of $\alpha_{\textrm{T}}$: it is the deviation
of the speed of tensors from the speed of light. The graviton gradient
terms are contained in the Ricci scalar of the spatial hypersurface
in unitary gauge, $^{(3)}R$. 

When evaluating the suitability of backgrounds, one must ensure that
all conditions \eqref{eq:scalarstab} and \eqref{eq:tensorstab} are
satisfied at all times of interest. When $\alpha_{\textrm{B}}\neq0$,
the region of phase space where $D=0$ represents a pressure singularity
\cite{Deffayet:2010qz} and no trajectory ever evolves across it \cite{Easson:2011zy}.
However, nothing in the background dynamics prevents a trajectory
with $c_{\text{s}}^{2}>0$ from crossing into an unstable region.
It is also possible that theories which violate the null energy condition
suffer from some sort of non-linear instability even when they do
not exhibit the instabilities described above \cite{Sawicki:2012pz}.

One may, in principle, also wish to constrain the viable models to
such trajectories where the speeds of sound are subluminal. Subluminality
is a necessary precondition for there to exist a standard Wilsonian
ultraviolet completion of the action \cite{Adams:2006sv}. However,
in general Horndeski models superluminalities always exist somewhere
in the phase space, especially in the presence of matter external
to the scalar \cite{Easson:2013bda}. There are not necessarily causal
paradoxes in theories with superluminality \cite{Babichev:2007dw,Burrage:2011cr},
but these theories must be ultraviolet completed in an alternative
way, e.g.\ via classicalisation \cite{Dvali:2010jz,Dvali:2010ns}.

\subsection{Linear Perturbations\label{sub:LinPerts}}

The value of the scalar field value is not an observable. We can reparameterise
the field by defining a new $\tilde{\phi}\equiv\tilde{\phi}(\phi)$
and all observables must remain unchanged. This means that focussing
on the scalar's value introduces unobservable redundancy into the
description. Instead, one can notice that the scalar-field gradient
forms a natural four-velocity,
\[
u_{\mu}\equiv-\frac{\partial_{\mu}\phi}{\sqrt{2X}}\,,
\]
which defines a comoving frame, provided that $\partial_{\mu}\phi$
be timelike, $X>0$. The meaning of $\phi$ then is one of a clock
and its gradient is the time direction for an observer at rest in
this frame. The perturbation of the scalar, when appropriately normalised,
can then be interpreted as a scalar potential for a peculiar velocity
field,
\begin{equation}
v_{X}\equiv-\frac{\delta\phi}{\dot{\phi}}\,,\label{eq:vx-def}
\end{equation}
where the overdot denotes the derivative with respect to the coordinate
time $t$. This velocity potential is then invariant under the reparameterisations
of $\phi$ and allows us to write down a much simpler set of perturbations
equations. Oscillating models, where the background field velocity
$\dot{\phi}$ crosses zero, cannot be directly described using our
formulation, since the singularities in \eqref{eq:vx-def} would have
to be appropriately accounted for.\\

The main result of this paper is the non-redundant formulation of
linear cosmological perturbation equations in a manner immediately
implementable in codes for calculating linear large-scale structure
such as CAMB \cite{Lewis:1999bs} or CLASS \cite{Lesgourgues:2011re,Blas:2010hb}.
We present the Einstein equations below in the user-friendly Newtonian
gauge and coordinate time, providing the conformal-time synchronous-gauge
version in Appendix~\ref{sec:SynchGauge}. We are not the first to
derive these equations (see refs~\cite{DeFelice:2011hq,Gubitosi:2012hu,Bloomfield:2012ff}
and section~\ref{sub:OmegasMean}), but we argue in this paper that
our choice of variables allows for the separation of different physical
effects and a natural limit to concordance cosmology. This allows
us to discuss the physics of general Horndeski theories with relative
ease in section~\ref{sec:Physics}. 

The effect of the presence of the Horndeski scalar is to introduce
the dynamical velocity potential $v_{X}$, describing the perturbation
of the scalar field through eq.~\eqref{eq:vx-def}. We keep these
terms on the left-hand side of the perturbed Einstein equations to
emphasise the gravity-like nature of this degree of freedom, resulting
from the universality of its coupling. In addition, the scalar's background
configuration changes the coefficients of the gravitational potentials
away from their standard values arising from the Einstein tensor.
On the right-hand-side of the Einstein equations lie the standard
contributions of all the matter sources in the cosmology: dark matter,
baryons, photons and neutrinos.\\

The Hamiltonian constraint (Einstein $(00)$ equation) takes the form:

\begin{align}
 & 3\left(2-\alpha_{\textrm{B}}\right)H\dot{\Phi}+\left(6-\alpha_{\textrm{K}}-6\alpha_{\textrm{B}}\right)H^{2}\Psi+\frac{2k^{2}\Phi}{a^{2}}\label{eq:Hamilt}\\
 & -\left(\alpha_{\textrm{K}}+3\alpha_{\textrm{B}}\right)H^{2}\dot{v}_{X}-\left[\alpha_{\textrm{B}}\frac{k^{2}}{a^{2}}-3\dot{H}\alpha_{\textrm{B}}+3\left(2\dot{H}+\tilde{\rho}_{\text{m}}+\tilde{p}_{\text{m}}\right)\right]Hv_{X}=-\tilde{\rho}_{\text{m}}\delta_{\text{m}}\,,\nonumber 
\end{align}
the momentum constraint (Einstein $(0i)$ equation)

\begin{equation}
2\dot{\Phi}+\left(2-\alpha_{\textrm{B}}\right)H\Psi-\alpha_{\textrm{B}}H\dot{v}_{X}-\left(2\dot{H}+\tilde{\rho}_{\text{m}}+\tilde{p}_{\text{m}}\right)v_{X}=-\left(\tilde{\rho}_{\text{m}}+\tilde{p}_{\text{m}}\right)v_{\text{m}}\,,\label{eq:Momentum}
\end{equation}
the anisotropy constraint (spatial traceless part of the Einstein
equations)
\begin{equation}
\Psi-\left(1+\alpha_{\textrm{T}}\right)\Phi-\left(\alpha_{\textrm{M}}-\alpha_{\textrm{T}}\right)Hv_{X}=\tilde{p}_{\text{m}}\pi_{\text{m}}\,,\label{eq:Aniso}
\end{equation}
and the pressure equation (spatial trace part of the Einstein equations)

\begin{align}
2\ddot{\Phi} & -\alpha_{\textrm{B}}H\ddot{v}_{X}+2\left(3+\alpha_{\textrm{M}}\right)H\dot{\Phi}+\left(2-\alpha_{\textrm{B}}\right)H\dot{\Psi}\label{eq:PresEq}\\
 & +\left[H^{2}\left(2-\alpha_{\textrm{B}}\right)\left(3+\alpha_{\textrm{M}}\right)-\left(\alpha_{\textrm{B}}H\right)^{.}+4\dot{H}-\left(2\dot{H}+\tilde{\rho}_{\text{m}}+\tilde{p}_{\text{m}}\right)\right]\Psi\nonumber \\
 & -\left[\left(2\dot{H}+\tilde{\rho}_{\text{m}}+\tilde{p}_{\text{m}}\right)+\left(\alpha_{\textrm{B}}H\right)^{.}+H^{2}\alpha_{\textrm{B}}\left(3+\alpha_{\textrm{M}}\right)\right]\dot{v}_{X}\nonumber \\
 & -\left[2\ddot{H}+2\dot{H}H\left(3+\alpha_{\textrm{M}}\right)+\dot{\tilde{p}}_{\text{m}}+\alpha_{\text{M}}H\tilde{p}_{\text{m}}\right]v_{X}=\delta p_{\text{m}}/M_{*}^{2}\,,\nonumber 
\end{align}
where we have kept the effective Planck mass on the right hand side
explicitly to stress that the matter-pressure perturbation refers
only to the matter sector and not any perturbations in the Planck
mass, which are already included on the left-hand side. 

Finally, the equation of motion for the scalar velocity potential
$v_{X}$,

\begin{align}
3H & \alpha_{\textrm{B}}\ddot{\Phi}+H^{2}\alpha_{\textrm{K}}\ddot{v}_{X}-3\left[\left(2\dot{H}+\tilde{\rho}_{\text{m}}+\tilde{p}_{\text{m}}\right)-H^{2}\alpha_{\textrm{B}}\left(3+\alpha_{\textrm{M}}\right)-\left(\alpha_{\textrm{B}}H\right)^{.}\right]\dot{\Phi}\label{eq:scalar}\\
 & +\left(\alpha_{\textrm{K}}+3\alpha_{\textrm{B}}\right)H^{2}\dot{\Psi}-2\left(\alpha_{\textrm{M}}-\alpha_{\textrm{T}}\right)H\frac{k^{2}}{a^{2}}\Phi-\alpha_{\textrm{B}}H\frac{k^{2}}{a^{2}}\Psi-\nonumber \\
 & -\left[3\left(2\dot{H}+\tilde{\rho}_{\text{m}}+\tilde{p}_{\text{m}}\right)-\dot{H}(2\alpha_{\textrm{K}}+9\alpha_{\textrm{B}})-\right.\nonumber \\
 & \left.\quad\quad-H\left(\dot{\alpha}_{\textrm{K}}+3\dot{\alpha}_{\textrm{B}}\right)-H^{2}\left(3+\alpha_{\textrm{M}}\right)\left(\alpha_{\textrm{K}}+3\alpha_{\textrm{B}}\right)\right]H\Psi+\nonumber \\
 & +\left[2\dot{H}\alpha_{\textrm{K}}+\dot{\alpha}_{\textrm{K}}H+H^{2}\alpha_{\textrm{K}}\left(3+\alpha_{\textrm{M}}\right)\right]H\dot{v}_{X}+H^{2}M^{2}v_{X}+\nonumber \\
 & +\left[-\left(2\dot{H}+\tilde{\rho}_{\text{m}}+\tilde{p}_{\text{m}}\right)+2H^{2}\left(\alpha_{\textrm{M}}-\alpha_{\textrm{T}}\right)+H^{2}\alpha_{\textrm{B}}\left(1+\alpha_{\textrm{M}}\right)+\left(\alpha_{\textrm{B}}H\right)^{.}\right]\frac{k^{2}}{a^{2}}v_{X}=0\,,\nonumber 
\end{align}
with

\begin{eqnarray}
H^{2}M^{2} & \equiv & 3\dot{H}\left[\dot{H}\left(2-\alpha_{\textrm{B}}\right)+\tilde{\rho}_{\text{m}}+\tilde{p}_{\text{m}}-H\dot{\alpha}_{\textrm{B}}\right]-3H\alpha_{\textrm{B}}\left[\ddot{H}+\dot{H}H\left(3+\alpha_{\textrm{M}}\right)\right]\,.\label{eq:Mass}
\end{eqnarray}
The system is completed by the standard evolution for the perturbations
of the combined matter sector --- $\delta_{\text{m}}$, $v_{\text{m}}$,
$\delta p_{\text{m}}$ and $\pi_{\text{m}}$ --- obtained through
the usual Boltzmann code.\\

The most important feature of this formulation of the equations is
that the evolution equations depend \emph{only }on the standard matter-sector
perturbation quantities, the background expansion history $H(t)$
(and its derivatives $\dot{H}$ and $\ddot{H}$), the background energy
density and pressure of the matter sector, and the values of the four
dimensionless functions of time $\alpha_{i}(t)$ which fully define
the properties of the dark-energy perturbations. All of these elements
are in general completely independent of each other and can take arbitrary
values. The values of the scalar field $\phi$ or the velocity $\dot{\phi}$
do not enter the any of the equations directly at all. This formulation
is purely in terms of physical quantities. \\

The expansion history $H(t)/H_{0}$ has been quite precisely observed
and is seen to match that of $\Lambda$CDM with parameter $\Omega_{\text{m}0}^{\Lambda\text{CDM}}$
quite closely (e.g.~\cite{Anderson:2013zyy,Betoule:2014frx}). One
can vary this expansion history somewhat at the price of making it
less compatible with distance measurements. Given that the matter
sector is understood at linear order (DM is essentially non-interacting
and follows geodesics, baryons and photons interact at high densities,
neutrinos do not interact) the evolution of the background evolution
of each of the components and perturbations is determined. However,
it is frequently not appreciated that the energy density of matter
today $\tilde{\rho}_{\text{m}0}$ is a free parameter that cannot
be determined from the expansion history without a prior knowledge
of the dark-energy evolution history \cite{Amendola:2012ky}: one
can keep $H(t)$ fixed to be e.g.~exactly $\Lambda$CDM while altering
the amount of dark matter and simultaneously appropriately changing
the evolution of the equation of state. Essentially, $\Omega_{\text{m}0}$
can only be determined from the growth of perturbations even in simple
dark-energy cases like quintessence.%
\footnote{Seen in this light, even cluster counts are a measure of the amplitude
of the gravitational potential and therefore a non-linear measure
of the perturbations and not of background evolution.%
}

Indeed, in our formulation, one can think of picking a value for $\tilde{\Omega}_{\text{m}0}\neq\Omega_{\text{m}0}^{\Lambda\text{CDM}}$
for a fixed $\Lambda$CDM expansion history as describing a dark-energy
model which tracks the energy density of the matter sector at early
times and therefore is a model of early dark energy (``EDE'') with
an effective parameterisation similar to ref.~\cite{Doran:2006kp}.
Such tracker models are tightly constrained by cosmic-microwave-background
data \cite{Reichardt:2011fv,Pettorino:2013ia,Ade:2013zuv}, but this
is assuming that the dark energy does not cluster. A dark energy that
tracks the background equation of state but clusters at all scales
during recombination and afterwards would act exactly like dark matter
and therefore is not constrained at all. Our formulation gives an
easy way to investigate the viability of such models.\\

We note that in the limit $\alpha_{i}\rightarrow0$ \emph{together
}with $2\dot{H}+\tilde{\rho}_{\text{m}}\rightarrow0$, the equations
(\ref{eq:Hamilt}-\ref{eq:PresEq}) reduce to the standard $\Lambda$CDM
Einstein equations, while the equation of motion for the scalar \eqref{eq:scalar}
identically vanishes, since the scalar degree of freedom is no longer
present. This means that, for sufficiently small deviations from this
limit, the growth of perturbations will be sufficiently close to that
of $\Lambda$CDM and this parameterisation has the concordance model
as its natural limit.

On the other hand, the limit $\alpha_{i}\rightarrow0$ but with a
non-$\Lambda$CDM expansion history is a little peculiar: eq.~\eqref{eq:scalar}
reduces to a constraint. the constraint structure of general relativity
is modified allowing for an arbitrary evolution of the background.
In fact, this limit is equivalent to the \emph{cuscuton }model of
refs \cite{Afshordi:2006ad,Afshordi:2007yx} which is a limit of the
k-\emph{essence} class of models where the sound speed is infinite.
It should be noted that such a limit is the proper way of modelling
perturbations within the context of a $w$CDM cosmology that preserves
general covariance of the full action.

Finally, as is well known (e.g.~\cite{Song:2006ej}), exactly $\Lambda$CDM
expansion history does not necessarily imply that perturbations grow
as in $\Lambda$CDM. As discussed above, this is obviously true for
models with EDE where $\tilde{\Omega}_{\text{m}0}\neq\Omega_{\text{m}0}^{\Lambda\text{CDM}}$.
However, even in the case $2\dot{H}+\tilde{\rho}_{\text{m}}=0$ the
Einstein equations differ by the arbitrary functions $\alpha_{i}$,
which alter the solutions on top of the concordance background and
can change growth rates.

\subsection{Connection to EFT\label{sub:OmegasMean}}

We have chosen the set of variables $\alpha_{i}$ related to the form
of the perturbations of the dark-energy EMT, since that is most simply
related to physical properties of the perturbations. However, our
approach is very much equivalent to the recent results obtained through
effective-field-theory methods \cite{Gubitosi:2012hu,Bloomfield:2012ff,Bloomfield:2013efa,Gleyzes:2013ooa,Piazza:2013pua}.
Since the Horndeski Lagrangian describes all theories with no more
than second derivatives in their equations of motion on the FRW background,
the two approaches differ by essentially a redefinition of variables
for this subclass of models. The EFT framework does
allow for an extension which involve third derivatives in the Einstein
equations, which cancel once the constraints are solved \cite{Gleyzes:2013ooa,Gleyzes:2014dya}.
In principle an extra property function would have to be added to
our formulation to describe such theories.

As we have explained above, our claim is that our variables are more
concretely connected to physical properties of the perturbations and
the energy-momentum tensor, whereas the EFT framework describes perturbations
through coefficients of particular operators permitted by the symmetries
of the Friedmann background. The benefit of our approach
is that it cleanly separates the properties of the background expansion
($\dot{H}$ and $\Omega_{\text{m}0}$) from those of the perturbations
($\alpha_{i}$), whereas the EFT approach mixes the two. Moreover,
as noted by ref.~\cite{Piazza:2013pua}, the EFT approach has some
redundancy, which when removed reduces to the same number of degrees
of freedom as we have in our minimal framework, five. We have provided
a mapping between the EFT operators and our $\alpha_{i}$ in table\textbf{~}\ref{tab:translations},
as well as definitions employed by other authors in the past to connect
all these approaches together.

\begin{table}
\begin{centering}
\begin{tabular}{llrrrrr}
\toprule 
\multicolumn{2}{l}{\textsf{\textbf{\scriptsize{}Variable Translations}}} & {\scriptsize{}$\boldsymbol{M_{*}^{2}}$} & \textsf{\textbf{\scriptsize{}$\boldsymbol{M_{*}^{2}H\alpha_{\textrm{M}}}$}} & \textsf{\textbf{\scriptsize{}$\boldsymbol{M_{*}^{2}H^{2}\alpha_{\textrm{K}}}$}} & \textsf{\textbf{\scriptsize{}$\boldsymbol{M_{*}^{2}H\alpha_{\textrm{B}}}$}} & \textsf{\textbf{\scriptsize{}$\boldsymbol{M_{*}^{2}\alpha_{\textrm{T}}}$}}\tabularnewline
\midrule
\midrule 
{\scriptsize{}Amendola et al. } & {\scriptsize{}\cite{Amendola:2012ky}} & \textsf{\textbf{\scriptsize{}$w_{1}$}} & \textsf{\textbf{\scriptsize{}$\dot{w}_{1}$}} & \textsf{\textbf{\scriptsize{}$\frac{2}{3}w_{3}+6Hw_{2}-6H^{2}w_{1}$}} & \textsf{\textbf{\scriptsize{}$-w_{2}+2Hw_{1}$}} & \textsf{\textbf{\scriptsize{}$w_{4}-w_{1}$}}\tabularnewline
\midrule 
{\scriptsize{}Bloomfield et al.} & {\scriptsize{}\cite{Bloomfield:2012ff,Bloomfield:2013efa}} & \textsf{\textbf{\scriptsize{}$m_{0}^{2}\Omega+\bar{M}_{2}^{2}$}} & \textsf{\textbf{\scriptsize{}$m_{0}^{2}\dot{\Omega}+\dot{\bar{M}}_{2}^{2}$}} & \textsf{\textbf{\scriptsize{}$2c+4M_{2}^{4}$}} & \textsf{\textbf{\scriptsize{}$-\bar{M}_{1}^{3}-m_{0}^{2}\dot{\Omega}$}} & \textsf{\textbf{\scriptsize{}$-\bar{M}_{2}^{2}$}}\tabularnewline
\midrule 
{\scriptsize{}De Felice et al. } & {\scriptsize{}\cite{DeFelice:2011hq}} & \textsf{\textbf{\scriptsize{}$\mathcal{G}_{T}$}} & \textsf{\textbf{\scriptsize{}$\dot{\mathcal{G}}_{T}$}} & \textsf{\textbf{\scriptsize{}$2\Sigma+12H\Theta-6H^{2}\mathcal{G}_{T}$}} & \textsf{\textbf{\scriptsize{}$-2\Theta+2H\mathcal{G}_{T}$}} & \textsf{\textbf{\scriptsize{}$\mathcal{F}_{T}-\mathcal{G}_{T}$}}\tabularnewline
\midrule 
{\scriptsize{}Gubitosi et al. } & {\scriptsize{}\cite{Gubitosi:2012hu,Gleyzes:2013ooa}} & \textsf{\textbf{\scriptsize{}$M_{*}^{2}f+2m_{4}^{2}$}} & \textsf{\textbf{\scriptsize{}$M_{*}^{2}\dot{f}+2\left(m_{4}^{2}\right)^{\cdot}$}} & \textsf{\textbf{\scriptsize{}$2c+4M_{2}^{4}$}} & \textsf{\textbf{\scriptsize{}$-m_{3}^{3}-M_{*}^{2}\dot{f}$}} & \textsf{\textbf{\scriptsize{}$-2m_{4}^{2}$}}\tabularnewline
\midrule 
{\scriptsize{}Piazza et al.} & {\scriptsize{}\cite{Piazza:2013pua}} & \textsf{\textbf{\scriptsize{}$M^{2}(1+\epsilon_{4})$}} & \textsf{\textbf{\scriptsize{}$\left(M^{2}(1+\epsilon_{4})\right)^{\cdot}$}} & \textsf{\textbf{\scriptsize{}$2M^{2}(\mathcal{C}+2\mu_{2}^{2})$}} & \textsf{\textbf{\scriptsize{}$-M^{2}(\mu_{3}+\mu)$}} & \textsf{\textbf{\scriptsize{}$-M^{2}\epsilon_{4}$}}\tabularnewline
\bottomrule
\end{tabular}
\par\end{centering}

\protect\caption{Translation table between the independent linear perturbation variables
$\alpha_{i}$ in our formulation and definitions used in other works.
Our approach is equivalent to the EFT methods, but removes the redundancy
in the description. For example, our $\alpha_{\textrm{K}}$ is a combination
of the EFT variables $c$ and $M_{2}^{4}$: they never appear separately
in the perturbations equations (\ref{eq:Hamilt}-\ref{eq:scalar})
or stability conditions (\ref{eq:scalarstab}-\ref{eq:tensorstab})
and therefore cannot be individually constrained, as already pointed
out in \eqref{eq:Hamilt}. We discuss the relation of our variables
to classes of dark energy models in table~\ref{tab:omegas} and their
impact on physics and therefore detectability in section~\ref{sec:Physics}.
\label{tab:translations}}
\end{table}

\section{Physics of Horndeski Linear Structure Formation\label{sec:Physics}}

We eliminate the scalar field $v_{X}$ using the constraints (\ref{eq:Hamilt}-\ref{eq:Aniso})
and the time derivative of eq.\ \eqref{eq:Aniso} to obtain a dynamical
equation for the potential $\Phi$ with a source driven by the matter
perturbations, 
\begin{equation}
\ddot{\Phi}+\frac{\beta_{1}\beta_{2}+\beta_{3}\alpha_{\textrm{B}}^{2}\frac{k^{2}}{a^{2}}}{\beta_{1}+\alpha_{\textrm{B}}^{2}\frac{k^{2}}{a^{2}}}\dot{\Phi}+\frac{\beta_{1}\beta_{4}+\beta_{1}\beta_{5}\frac{k^{2}}{a^{2}}+c_{\text{s}}^{2}\alpha_{\textrm{B}}^{2}\frac{k^{4}}{a^{4}}}{\beta_{1}+\alpha_{\textrm{B}}^{2}\frac{k^{2}}{a^{2}}}\Phi=-\frac{1}{2}\tilde{\rho}_{\text{m}}\frac{\beta_{1}\beta_{6}+\beta_{7}\alpha_{\textrm{B}}^{2}\frac{k^{2}}{a^{2}}}{\beta_{1}+\alpha_{\textrm{B}}^{2}\frac{k^{2}}{a^{2}}}\delta_{\text{m}}\,,\label{eq:dynamical}
\end{equation}
where the functions $\beta_{i}\equiv\beta_{i}\left(t\right)$ are
all functions of the property functions $\alpha_{i}$ and the background
$H(t)$, defined in appendix \ref{sec:Scale-Dependence-in}. We have
neglected the matter velocity term since for the purposes of this
section we will assume it that the matter is dust and therefore the
velocities are irrelevant inside the cosmological horizon. We can
obtain the Newtonian potential $\Psi$ through a version of the constraint
\eqref{eq:Aniso} with the scalar field eliminated,
\begin{align}
\alpha_{\textrm{B}}^{2}\frac{k^{2}}{a^{2}} & \left[\Psi-\Phi\left(1+\alpha_{\textrm{T}}+\frac{2\left(\alpha_{\textrm{M}}-\alpha_{\textrm{T}}\right)}{\alpha_{\textrm{B}}}\right)\right]+\label{eq:anisotropy}\\
+ & \beta_{1}\left[\Psi-\Phi\left(1+\alpha_{\textrm{T}}\right)\left(1-\frac{2DH^{2}\left(\alpha_{\textrm{M}}-\alpha_{\textrm{T}}\right)}{\beta_{1}}\right)\right]=\left(\alpha_{\textrm{M}}-\alpha_{\textrm{T}}\right)\left[\alpha_{\textrm{B}}\tilde{\rho}_{\textrm{m}}\delta_{\textrm{m}}-2HD\dot{\Phi}\right]\,.\nonumber 
\end{align}
Augmented with the standard evolution equations for dust perturbations
\begin{equation}
\dot{\delta}_{\text{m}}-\frac{k^{2}}{a^{2}}v_{\text{m}}=3\dot{\Phi}\,,\qquad\dot{v}_{\text{m}}=-\Psi\,,\label{eq:MatterCons}
\end{equation}
eqs\ \eqref{eq:dynamical} and \eqref{eq:anisotropy} form the complete
dynamical system.\\

The equations presented in section\ \ref{sub:LinPerts} describe
in full generality the evolution of first-order scalar perturbations
in models within the scope defined in section~\ref{sec:Scope}, and
they are in the natural form for implementation in numerical codes
such as CAMB \cite{Lewis:1999bs} and CLASS \cite{Lesgourgues:2011re,Blas:2011rf}.
However, for the purpose of interpreting the physics, it is helpful
to eliminate the scalar field $v_{X}$ and express all of the dynamics
in terms of the variables familiar from the standard $\Lambda$CDM
case.

We have written the equations \eqref{eq:dynamical} and \eqref{eq:anisotropy}
in a particular manner to make explicit the existence of a new transition
scale in the behaviour of the dynamics, which we will call the \emph{braiding
scale} $k_{\text{B}}$. Whenever there is any kinetic braiding, $\alpha_{\textrm{B}}\neq0$,
then the response of the gravitational potentials will transition
between two separate regimes at the scale 
\begin{equation}
\frac{k_{\text{B}}^{2}}{a^{2}H^{2}}\equiv\frac{\beta_{1}}{\alpha_{\textrm{B}}^{2}H^{2}}=\frac{D}{\alpha_{\textrm{B}}^{2}}\left[(1-\tilde{\Omega}_{\text{m}})(1+w_{X})+2\left(\alpha_{\textrm{M}}-\alpha_{\textrm{T}}\right)\right]+\frac{9}{2}\tilde{\Omega}_{\text{m}}\,.\label{eq:TransitionScale}
\end{equation}
During matter domination, the braiding scale --- if it at all exists,
$\alpha_{\textrm{B}}\neq0$ --- typically lies at the cosmological
horizon, but can lie inside when
\[
\frac{\alpha_{\textrm{K}}}{\alpha_{\textrm{B}}^{2}}(1-\tilde{\Omega}_{\text{m}})(1+w_{X})\gg1\,.
\]
The same braiding scale $k_{\text{B}}$ appears in both the dynamical
equation \eqref{eq:dynamical} and in the anisotropy constraint \eqref{eq:anisotropy}
and was discussed for the first time in ref.~\cite{Sawicki:2012re}
in the context of a small subclass of the models being considered
here.%
\footnote{It was named the imperfect scale in ref.~\cite{Sawicki:2012re},
since the anisotropic stress vanished outside it. This was the result
of a Brans-Dicke-type non-minimal coupling considered, but it is not
the general behaviour, as shown in section~\ref{sub:Example:-No-braiding}.%
} If $k_{\text{B}}$ lies subhorizon, this is evidence of a hierarchy
between the values of $\alpha_{\textrm{K}}$ and $\alpha_{\textrm{B}}$.
In archetypal modified-gravity models such as $f(R)$ and $f(G)$,
$\alpha_{\textrm{K}}=0$ and therefore $k_{\text{B}}$ always lies
either close to or outside the cosmological horizon. The whole subhorizon
regime always exhibits braiding in these models (in the linear regime).\\

When considering the observable impact, it is usual to compress any
modifications from the concordance case into an effective Newton's
constant and the slip parameter. Since there are two potentials, two
effective Newton's constants can be defined,
\begin{equation}
Y\equiv-\frac{2k^{2}\Psi}{a^{2}\tilde{\rho}_{\textrm{m}}\delta_{\textrm{m}}}\,,\qquad Z\equiv-\frac{2k^{2}\Phi}{a^{2}\tilde{\rho}_{\textrm{m}}\delta_{\textrm{m}}}\,.\label{eq:geff}
\end{equation}
One usually discusses $Y$, since this is the term that enters directly
in the equations for growth rate for matter perturbations (see e.g.~\cite{Amendola:2007rr}).
However, it is actually $\Phi$ that is the dynamical variable (related
to the spatial curvature perturbation) and it turns out to generically
have a simpler behaviour. Thus $Z$ should be considered more amenable
to parameterisation in the classes of models considered here. Since
it is $Z$ that appears in the Hamiltonian constraint, eq.~\eqref{eq:Hamilt},
it is a deviation of $Z$ from its subhorizon GR value of 1 that signifies
that dark energy clusters.

An honestly model-independent observable that can be \emph{measured
}by comparing weak lensing and redshift-space distortions is the slip
parameter describing the anisotropic stress, 
\begin{equation}
\bar{\eta}\equiv\frac{2\Psi}{\Psi+\Phi}\,.\label{eq:slip}
\end{equation}
We have chosen a non-standard definition of this ratio to reflect
the fact that projected measurement errors from Euclid are minimised
for this particular combination and could be as low as a few percent
under certain assumptions \cite{Amendola:2013qna}. A parameterisation
of one of the effective Newton's constants \eqref{eq:geff} and the
slip parameter \eqref{eq:slip} is sufficient to describe the dynamics
of the matter sector and therefore to calculate the observables. These
can be translated to observables in more frequent use by the community
through
\[
Y=\frac{\bar{\eta}}{2-\bar{\eta}}Z\qquad\Sigma=\frac{2Z}{2-\bar{\eta}}\,.
\]

We devote the remainder of this section to describing the typical
behaviour of these variables in the class of models within our scope
to build an understanding of what can be expected in models which
are fully consistent rather than purely phenomenological. First, one
can ignore all the scale dependence and seek the extreme quasi-static
limit of the dynamics, $k\rightarrow\infty$, where
\begin{align}
Z_{\text{QS}} & =1+\frac{\alpha_{\textrm{B}}^{2}\left(1+\alpha_{\textrm{T}}\right)+2\alpha_{\textrm{B}}\left(\alpha_{\textrm{M}}-\alpha_{\textrm{T}}\right)}{2Dc_{\text{s}}^{2}}\,,\label{eq:EQS}\\
\bar{\eta}_{\text{QS}} & =1+\frac{2\left(\beta_{7}-c_{\text{s}}^{2}\right)\left(\alpha_{\textrm{M}}-\alpha_{\textrm{T}}\right)+\beta_{7}\alpha_{\textrm{B}}\alpha_{\textrm{T}}}{2\left(\beta_{7}-c_{\text{s}}^{2}\right)\left(\alpha_{\textrm{M}}-\alpha_{\textrm{T}}\right)+\beta_{7}\left(2\alpha_{\textrm{B}}+\alpha_{\textrm{B}}\alpha_{\textrm{T}}\right)}\,,\qquad\alpha_{\textrm{B}}\neq0\,.\nonumber 
\end{align}
We note here that the dark energy clusters at small scales ($Z\neq1$)\emph{
only }if there is braiding, $\alpha_{\textrm{B}}\neq0$. Thus a detection
of clustering of dark energy is unambiguous evidence of the presence
of kinetic mixing of the scalar and graviton. Secondly, the limit
for $\bar{\eta}$ taken in \eqref{eq:EQS} does not exist when there
is no braiding (see section~\ref{sub:Example:-No-braiding} for an
account of what happens when $\alpha_{\textrm{B}}=0$). 

In the following subsections, we discuss in more detail two limiting
cases which put the above observation on firmer footing: 
\begin{itemize}
\item No braiding, $\alpha_{\textrm{B}}=0$; the braiding scale $k_{\text{B}}$
does not exist. The sound speed provides the only scale in the problem.
Inside the Jeans scale, dark energy does not cluster, $Z=1$. Anisotropic
stress can be non-vanishing, with a slip parameter constant as a function
of scale inside the Jeans length.
\item Negligible standard kinetic term, $\alpha_{\textrm{K}}\ll\alpha_{\textrm{B}}^{2}$;
the braiding scale $k_{\text{B}}$ is superhorizon. There is a single
transition scale determined by the Compton mass of the scalar. DE
clusters only inside the Compton scale. The slip parameter interpolates
between two values with the transition also occurring at the Compton
scale.
\end{itemize}
In principle, the fully general model will contain another scale (the
braiding scale $k_{\text{B}}$) across which the behaviour transitions
between these two behaviours (see the behaviour in ref.~\cite{Sawicki:2012re}).
We defer a detailed description of the phenomenology to a numerical
analysis, but stress that in general such properties as anisotropic
stress and DE clustering are scale-independent only for very particular
subclasses of dark-energy models.

\subsection{Example: No Braiding\label{sub:Example:-No-braiding}}

Here we assume that there is no braiding: the graviton and the scalar
do not mix kinetically at any scale, $\alpha_{\textrm{B}}=0$. This
set of models can be thought of as a generalisation of perfect-fluid
models to include anisotropic stress. The new scale dependence in
the coefficients of eqs.~(\ref{eq:dynamical}-\ref{eq:anisotropy})
disappears and we remain with

\begin{equation}
\ddot{\Phi}+H\left(4+\alpha_{\textrm{M}}+3\Upsilon\right)\dot{\Phi}+\left(\beta_{4}+\frac{c_{\text{s}}^{2}k^{2}}{a^{2}}\right)\Phi=-\frac{1}{2}c_{\text{s}}^{2}\tilde{\rho}_{\textrm{m}}\delta_{\textrm{m}}\,,\label{eq:nobdynamical}
\end{equation}
for the dynamical equation. The variable $\Upsilon$ is defined in
eq.~\eqref{eq:Upsilon}; it can be thought of as a generalisation
of the adiabatic sound speed for cases where the dark energy is not
a perfect fluid. The structure of eq.~\eqref{eq:nobdynamical} is
fundamentally unchanged from the case of the perfect fluid ($\alpha_{\textrm{M}}=\alpha_{\textrm{T}}=0$).
In particular, inside the Jeans length the effective Newton's constant
\begin{equation}
Z\simeq1\qquad\text{for}\qquad\frac{k^{2}}{a^{2}}\gg\beta_{4}/c_{\text{s}}^{2}\label{eq:NoGMod}
\end{equation}
no matter what other modifications are present. Since $\beta_{4}\sim H^{2}$,
this range of validity of this quasi-static approximation is essentially
determined by the sound speed of the dark energy.

Despite the fact that there is no modification in \eqref{eq:NoGMod},
and therefore --- just as in the case of a perfect-fluid dark energy
--- it does not cluster, growth rates can be affected, since this
sort of dark energy can carry anisotropic stress. The anisotropy constraint
reduces to

\begin{equation}
Hc_{\text{s}}^{2}\alpha_{\textrm{K}}\Psi+H\left[c_{\text{s}}^{2}\alpha_{\textrm{K}}+2\left(\alpha_{\textrm{M}}-\alpha_{\textrm{T}}\right)\right]\left(1+\alpha_{\textrm{T}}\right)\Phi=-2\left(\alpha_{\textrm{M}}-\alpha_{\textrm{T}}\right)\dot{\Phi}\,,\label{eq:nobanisotropy}
\end{equation}
i.e.~it also loses any scale dependence of the coefficients as well
as its dependence on $\delta_{\text{m}}$. The only scale dependence
that can appear in the slip parameter is through $\dot{\Phi}$, which
cannot be neglected here, especially if $c_{\text{s}}^{2}\ll1$.%
\footnote{This is an example of the failure of the quasi-static approximation,
see section \ref{sub:QS}.%
} The scale-independent result valid in the regime of \eqref{eq:NoGMod},
is

\begin{equation}
\bar{\eta}=1+\frac{c_{S}{}^{2}\alpha_{\textrm{K}}\alpha_{\textrm{T}}+2\left(\alpha_{\textrm{M}}-\alpha_{\textrm{T}}\right)^{2}}{c_{S}{}^{2}\alpha_{\textrm{K}}\left(2+\alpha_{\textrm{T}}\right)+2\left(\alpha_{\textrm{M}}-\alpha_{\textrm{T}}\right)^{2}}\,.\label{eq:nobeta}
\end{equation}

\subsection{Example: Negligible Standard Kinetic Term\label{sub:Example:-Negligible-standard}}

In this section, we assume that the kinetic term of the scalar mode
is mainly produced through mixing with the graviton rather than directly,
$\alpha_{\textrm{B}}^{2}\gg\alpha_{\textrm{K}}$. For the sake of
simplicity of presentation, we also choose $\alpha_{\textrm{T}}=0$.
These kind of models can be considered as a generalization of the
$f(R)$ theories, since the braiding $\alpha_{\textrm{B}}$ and the
Planck-mass run rate $\alpha_{\textrm{M}}$ are uncorrelated here
(see table\ \ref{tab:omegas}). More generally, we are essentially
describing the behaviour of other modified gravity models such as
$f(G)$ and models with large braiding such as Imperfect Dark Energy
of ref.~\cite{Deffayet:2010qz}.

The positivity of the sound speed \eqref{eq:scalarstab} provides
an upper limit to the magnitude of the braiding $\alpha_{\textrm{B}}$.
We will take it to be

\begin{equation}
\alpha_{\textrm{B}}\lesssim\tilde{\Omega}_{X}\,.
\end{equation}
Under this assumption, the braiding scale \eqref{eq:TransitionScale}
always lies superhorizon, 
\begin{equation}
\frac{k_{\text{B}}^{2}}{a^{2}}\equiv\frac{\beta_{1}}{\alpha_{\textrm{B}}^{2}}\lesssim\mathcal{O}(H^{2})\,.\label{eq:NoKbraidingscale}
\end{equation}
and therefore inside the observable domain the dark energy scalar
is always kinetically mixed with the graviton. In this subhorizon
regime, the evolution equation \eqref{eq:dynamical} reduces to

\begin{eqnarray}
\ddot{\Phi}+\left(3+\alpha_{\textrm{M}}\right)H\dot{\Phi} & + & \left(\frac{\beta_{1}\beta_{5}}{\alpha_{\textrm{B}}^{2}}+c_{\text{s}}^{2}\frac{k^{2}}{a^{2}}\right)\Phi\simeq-\frac{1}{2}\tilde{\rho}_{\text{m}}\left(\frac{\beta_{1}\beta_{6}}{\alpha_{\textrm{B}}^{2}\frac{k^{2}}{a^{2}}}+\beta_{7}\right)\delta_{\text{m}}\,.\label{eq:nokpoisson}
\end{eqnarray}
Yet again, a scale is present in this equation
\[
\frac{k_{\text{C}}^{2}}{a^{2}}\equiv\frac{\beta_{1}\beta_{5}}{\alpha_{\textrm{B}}^{2}c_{\text{s}}^{2}}
\]
which in the context of $f(R)$ gravity is called the Compton mass
scale \cite{DeFelice:2010aj}. On either side of this scale, in the
quasi-static limit, we have 
\begin{align}
Z= & \frac{\beta_{6}}{\beta_{5}} & aH\ll k\ll k_{\text{C}} & \,,\\
Z= & \frac{\beta_{7}}{c_{\text{s}}^{2}}=1+\frac{1}{3c_{\text{s}}^{2}}+\frac{2\alpha_{\textrm{M}}}{3\alpha_{\textrm{B}}c_{\text{s}}^{2}} & k\gg k_{\text{C}} & \,.\nonumber 
\end{align}
The requirement that $Z$ remain of the order of 1 in order to not
catastrophically affect structure formation ensures that the coefficients
of $\Phi$ and $\delta_{\text{m}}$ in eq.~\eqref{eq:nokpoisson}
both transition in their behaviour around the same scale. Thus, despite
the rather peculiar $k$-dependence, the behaviour of the braided
models is to interpolate between two different values of the effective
Newton's constant $Z$ across the Compton scale. It is typical for
$\beta_{5}\gg1$ before dark-energy domination, ensuring that most
of the scales relevant to structure formation observations are super-Compton.
We also note that $Z$ is naturally close to 1 when the parameters
$\alpha_{i}$ are not tuned to create large hierarchies. 

The only scale present in the coefficients of the anisotropy equation
\eqref{eq:anisotropy} is the scale \eqref{eq:NoKbraidingscale} which
is superhorizon at all times. Therefore the only scale dependence
in the slip parameter is a result of the scale dependence in $Z$,
thus we obtain
\begin{align}
\bar{\eta} & \equiv1-\frac{\frac{\alpha_{\textrm{M}}}{\alpha_{\textrm{B}}}\left(\alpha_{\textrm{M}}+2\Upsilon\right)}{1+\Upsilon-\frac{3}{2}\alpha_{\textrm{B}}c_{S}{}^{2}-\frac{\alpha_{\textrm{M}}}{\alpha_{\textrm{B}}}\left(\alpha_{\textrm{M}}+2\Upsilon\right)}\qquad & aH\ll k\ll k_{\text{C}}\\
\bar{\eta} & \equiv1+\frac{\frac{\alpha_{\textrm{M}}}{\alpha_{\textrm{B}}}\left(1+2\frac{\alpha_{\textrm{M}}}{\alpha_{\textrm{B}}}\right)}{3c_{S}{}^{2}+\left(1+\frac{\alpha_{\textrm{M}}}{\alpha_{\textrm{B}}}\right)\left(1+2\frac{\alpha_{\textrm{M}}}{\alpha_{\textrm{B}}}\right)} & k\gg k_{\text{C}}\nonumber 
\end{align}
where $\Upsilon$ is defined in eq.~\eqref{eq:Upsilon} and is a
variable that reduces to the adiabatic sound speed in the case of
a perfect fluid. Thus in general the anisotropic stress can be non-vanishing
both inside and outside the Compton scale. 

A natural limit of this class of models is $f(R)$ gravity for which
$\alpha_{\textrm{M}}=-\alpha_{\textrm{B}}$. It can easily be checked
that at super-Compton scales both $Z$ and $\bar{\eta}$ are very
close to their $\Lambda$CDM values of 1 (with corrections of order
$\alpha_{\textrm{M}}$), while inside the Compton scale we have the
standard result $Z=2/3$ and $\bar{\eta}=4/3$.

\subsection{Comment on Quasi-Static Limit\label{sub:QS}}

Typically when discussing modified-gravity models, the quasi-static
approximation (QS) is used to obtain the effective Newton's constant
$Z$ and the slip parameter $\bar{\eta}$. This involves neglecting
all terms time derivatives in eq.~\eqref{eq:scalar}, turning a dynamical
equation into a constraint between the value of the velocity potential
$v_{X}$ and the gravitational potential. This result is then used
in the Hamiltonian constraint \eqref{eq:Hamilt} and the anisotropy
constraint \eqref{eq:Aniso} under a similar approximation to obtain
$Z$ and $\bar{\eta}$. Does this procedure give the same result as
the quasi-static limit of eq.~\eqref{eq:dynamical} which was obtained
without such approximations?

It turns out that the two results are identical in the $k\rightarrow\infty$
limit \eqref{eq:EQS}. However the term subleading in $k^{2}$ are
affected. The reason for this is that both equation \eqref{eq:Momentum}
and \eqref{eq:PresEq} do not have any scale dependence. As a result
eliminating, the time derivatives of $v_{X}$ in eq.~\eqref{eq:scalar}
does not affect any of the $k^{2}$ terms. However the other terms
\emph{are} changed, leading to non-negligible differences at larger
scales. One must therefore be careful about how the quasi-static limit
is taken when calculating e.g.~the Compton mass, as in section~\ref{sub:Example:-Negligible-standard}.\\

A separate issue is whether the QS limit is at all a good approximation
to the full dynamics at small-enough scales. In reality, taking the
QS approximation is equivalent to turning a full degree of freedom
into a constraint. In principle the dynamics of the late universe
are described by the coupled system of equations \eqref{eq:dynamical}
and conservation for the matter EMT \eqref{eq:MatterCons}. One should
investigate the normal modes of this coupled system and ask what their
behaviour is. It may well be that an instability exists which the
quasi-static limit would hide.

In particular, the QS approximation removes from consideration the
oscillating modes which solve the homogeneous version of eq.~\eqref{eq:dynamical}.
Whether those decay or grow compared to the solutions obtained in
the QS limit depends, among others, on the friction term in \eqref{eq:dynamical},
i.e.~on the values of $\beta_{2}$ and $\beta_{3}$. If all the $\alpha_{i}\lesssim1-\tilde{\Omega}_{\text{m}}$,
then $\beta_{2}\sim4H+\mathcal{O}(\alpha_{i}H)$ and $\beta_{3}\sim3H+\mathcal{O}(\alpha_{i}H)$
and therefore there is no large difference between the friction terms
on any scales. Inside the braiding scale the homogeneous mode may
be marginally more unstable, but probably this is not something particularly
dangerous for most models. The details do however depend on the behaviour
of the mass term and should be studied more extensively.

\subsection{Constraints beyond Large-Scale Structure}

In addition to the effect on cosmological large-scale structure driven
by the modifications in section~\ref{sub:LinPerts}, the realised
dynamical dark energy mechanism must satisfy additional constraints
(see e.g.~refs~\cite{Uzan:2010pm,Will:2014kxa} and references therein)
\begin{enumerate}
\item Bing-Bang Nucleosynthesis constrains the expansion rate to be within
approximately 10\% of the standard one imputed using local measurements
of Newton's constant. This is a restriction on the evolution history
of the Planck mass $M_{*}$.
\item Shapiro-time-delay tests in the Solar System constrain the parameterised
post-Newtonian parameter $\left|\gamma-1\right|<10^{-5}$. This is
a restriction on the anisotropic stress present around the solar solution,
i.e.~the slip parameter $\bar{\eta}$.
\item Binary-pulsar orbits decay in a manner consistent with general relativity.
A new gravitationally coupled scalar is a new channel for radiation
and therefore accelerated orbit evolution \cite{Yagi:2013ava}.
\item If gravity is \emph{slower }than the speed of light, Čerenkov radiation
into gravitons would be produced by particles moving sufficiently
quickly, such as cosmic rays. 
\end{enumerate}
One must be careful in how one interprets these constraints in the
context of dynamical dark energy. The linear perturbation equations
(\ref{eq:Hamilt}-\ref{eq:scalar}) must remain valid whenever the
background has all the symmetries of an FRW universe, i.e.~well inside
any localised distribution of matter which is sufficiently homogeneous
and isotropic. However, it is not necessarily true that the local
values of the parameters $\alpha_{i}$ are the same as those for the
cosmology at large scales. As a result of screening around localised
objects, a transition region can appear where eqs~(\ref{eq:Hamilt}-\ref{eq:scalar})
are no longer valid and the solution interpolates between the cosmological
and the local values of the $\alpha_{i}$ (see e.g.~\cite{Brax:2012jr,Babichev:2013usa}
and references therein for a discussion of screening mechanisms). 

The constraints (1) and (2) above should therefore be seen as constraints
on the local or \emph{Solar System }values of the $\alpha_{i}$ and
not on their cosmological values. In particular, constraint (1) in
our context should be interpreted as 
\[
\left|\Omega_{\text{m}}(t_{\text{BBN}})Y_{\text{SS}}\frac{M_{*}^{2}(t_{\text{BBN}})}{M_{*,\text{SS}}^{2}}-1\right|\lesssim10\%\,.
\]
No measurement of the pure Planck mass exists, but rather the force
felt by masses in the Solar-System is a sum of both the gravitational
and the scalar force and is described by the local values of the effective
Newton's constant $Y_{SS}$ and the the Solar-System value of $M_{*,\text{SS}}$,
which may not be the same as the one at large distances today. On
the other hand, the constraints on the expansion rate during BBN are
sensitive to both the value of the Planck mass and on the presence
of any early dark energy at that time. 

Similarly, constraint (2) is sensitive to the Solar System value of
the slip parameter, i.e.
\[
\left|\bar{\eta}_{\text{SS}}-1\right|=\left|\frac{\gamma_{\text{SS}}-1}{2}\right|\lesssim10^{-5}
\]
which is related to the cosmological one through the screening mechanism,
just as in ref.~\cite{Hu:2007nk}. The knowledge of the screening
mechanism is required to answer whether screening can happen in the
Solar System, but this requires the study of higher-order perturbation
theory and cannot be answered within the context of the linear one.
Only in models where no screening occurs, the Solar-System values
are the same as the cosmological ones and the constraints above apply
directly.

Binary-pulsar orbits in principle constrain our models at various
locations in the Galaxy. However, since neutron stars are much more
compact than the Sun, the level of screening is likely to be much
stronger than in the Solar System and therefore the constraints they
provide are related to even denser environments than those of the
Solar System.

Finally, constraint (4) is relevant both to the local tests but also
the wider cosmological scales. The requirement that no gravitational
Čerenkov radiation be emitted is a lower bound on the speed of gravity
\[
\alpha_{\textrm{T}}\gtrsim-10^{-15}\,,
\]
but no such upper limit exists on cosmological scales \cite{Moore:2001bv,Elliott:2005va,Kimura:2011qn}.

\section{Discussion and Conclusions\label{sec:concl}}

In this paper, we have presented a turnkey solution for the study
of linear cosmological perturbations in general dark-energy and modified-gravity
models described by the Horndeski Lagrangian, with a formulation focussing
on non-redundant variables which lead to physical effects. Our approach
is essentially equivalent to the recent work within the framework
of effective field theory, amounting to a redefinition of variables
in this context. However, we make explicit the variables which independently
affect the behaviour of perturbations and which are constrainable
by observations. The dark-energy/gravity sector of an arbitrary general
Horndeski model can be described by five functions of time and one
constant:
\begin{enumerate}
\item A completely arbitrary dimensionless background expansion history
$H(t)/H_{0}$, which is all that observations of distances, using
supernovae, BAOs can map out (in principle, spatial curvature $\Omega_{k0}$
can also be allowed as an extra constant parameter and unambiguously
measured using longitudinal BAO \cite{Amendola:2012ky}).
\item The value of fractional matter density $\Omega_{\text{m}0}$. This
is a constant not determined by measurements of the background expansion
history since one can always change the amount of dark matter and
replace it with an appropriately evolving dark energy. More appropriately,
one should think of $\Omega_{\text{m}0}$ as being a parameter which
can only be measured by observing the evolution of cosmological perturbations,
whether in the cosmic microwave background or at late times.
\item Four arbitrary dimensionless property functions $\alpha_{i}(t)$ defining
the effect of the dark energy model on the evolution of linear perturbations
and independent from the above and each other in the most general
case.
\end{enumerate}
In addition, to convert these prediction to physical units, a value
of $H_{0}$, the Hubble parameter today, is necessary. Local measurements
of supernovae are affected by cosmic variance and therefore not quite
true measurements of the averaged $H_{0}$, especially in the sense
independent of the dark-energy model \cite{Marra:2013rba,Ben-Dayan:2014swa}.
Alternatively, measurements of age differences of red galaxies can
be used, as proposed in refs~\cite{Jimenez:2001gg,Simon:2004tf}
(see also refs~\cite{Stern:2009ep,Moresco:2010wh,Bellini:2013hea}).

All the subclasses of dark-energy models in our scope are described
as lying in a subspace of the four dimensions spanned by $\alpha_{i}$.
For example, quintessence models offer no freedom whatsoever once
(1) and (2) have been fixed; each of perfect fluid (k-\emph{essence}),
$f(R)$ and $f(G)$ models are completely specified by (1) and (2)
and then a one-dimensional subspace relating the functions $\alpha_{i}$
to each other. We have specified the restrictions on the full freedom
that various popular subclasses of dark energy models imply in table~\ref{tab:omegas}.\\

Even though there exists a Horndeski model which would describe any
background for any choice of the functions $\alpha_{i}$, not all
such configurations are stable. For any particular choice of background
expansion history and $\Omega_{\text{m}0}$, only certain ranges of
$\alpha_{i}$ are permitted, as implied by the inequalities given
in section~\ref{sub:backStab}. The background is unstable to scalar
modes whenever there are ghosts, $\alpha_{\textrm{K}}<-\frac{3}{2}\alpha_{\textrm{B}}^{2}$,
or the sound speed squared $c_{\text{s}}^{2}$ is smaller than zero.
In particular, this means that any perfect-fluid model ($\alpha_{\textrm{M}}=\alpha_{\textrm{B}}=\alpha_{\textrm{T}}=0$)
is unstable whenever the DE has $w_{X}<-1$. In addition, one must
pay attention to the tensor modes: whenever the Planck mass $M_{*}^{2}$
is evolving, it cannot be allowed to become negative in the past.
The stability conditions specified here when tested for over the whole
relevant evolution history should be considered a prerequisite for
the acceptance of a particular linear solution, even if linear codes
would seemingly allow an instability within the region of interest.\\

The evolution of perturbations is completely described by the set
of modified Einstein equations (\ref{eq:Hamilt}-\ref{eq:scalar})
(or, alternatively, their synchronous-gauge versions in appendix~\ref{sec:SynchGauge}).
All the effects of dynamical dark energy are included as modifications
to the left-hand side of the equations, with the contributions from
the matter EMT on the right-hand side standard in all but one way:
if the Planck mass is evolving ($\alpha_{\textrm{M}}\neq0$), the
energy density of \emph{none} of the matter species is conserved,
but must be tracked according to eq.~\eqref{eq:Conservation}. 

In order to access the physics of these models, we have presented
the full evolution equations for the gravitational potentials with
the unobservable scalar degree of freedom eliminated in section~\ref{sec:Physics}.
We have demonstrated the following:
\begin{itemize}
\item The effect of dark energy on the effective Newton's constant $Z$
is simpler than on $Y$ and parameterisations should in general focus
on $Z$ and $\bar{\eta}$. 
\item There is a braiding scale $k_{\text{B}}$ driven by the competition
between the kineticity $\alpha_{\textrm{K}}$ and kinetic braiding
$\alpha_{\textrm{B}}$, eq.\textbf{~}\eqref{eq:TransitionScale}.\textbf{
}This scale may be subhorizon and separates two behaviours in $Z$:

\begin{itemize}
\item \emph{No braiding, $\alpha_{\textrm{B}}=0$:} perfect-fluid-like behaviour
for $Z$: $Z=1$ inside Jeans length, and usual gravitational instability
outside it.
\item \emph{Negligible kineticity, $\alpha_{\textrm{K}}$ small:} Dark energy
clusters, $Z\neq1$. Transition between two regimes at scales related
to Compton mass of the scalar.
\end{itemize}
\item Anisotropic stress, $\bar{\eta}\neq1$, can only be present at linear
order whenever the tensor modes are non-minimally coupled, i.e.~either
$\alpha_{\textrm{M}}\neq0$ or $\alpha_{\textrm{T}}\neq0$. Its presence
is completely independent of values of $\alpha_{\textrm{K}}$ and
$\alpha_{\textrm{B}}$, but its value depends on all property functions.
\end{itemize}
If no deviation from the $\Lambda$CDM expansion history is detected
and the growth of perturbations is consistent with $\alpha_{i}=0$
up to some level of precision, then one should think of this result
as a statement that dynamical dark energy is responsible for at most
a small part of the acceleration mechanism. This sort of argument
was put forward in the case of models with chameleon screening in
ref.~\cite{Wang:2008zh}. The remainder of the acceleration can then
be considered to be driven by a pure non-dynamical cosmological constant.
For the $\Lambda$CDM expansion history, there essentially always
exists a family of models which interpolate between the cosmological
constant as the only source of acceleration and models with $\alpha_{i}\approx\mathcal{O}\left(1-\tilde{\Omega}_{\text{m}}\right)$
which are fully dynamical. \\

The formulation we have proposed is limited in scope to universally
coupled dark-energy models. However, it can be extended to also cover
non-universal couplings with the addition of new parameters (some
relative density of the non-universally coupled specie and the rate
of evolution of the coupling). In principle, only the right-hand side
of the Einstein equations and the (non)-conservation equation for
the differently coupled species need be modified. Adding additional
scalar degrees of freedom to the dark energy is a much more complicated
proposition. No full description of such a multi-scalar model exists
(the multi-field Horndeski action conjectured in \cite{Padilla:2012dx}
does not seem to contain all possible terms \cite{Kobayashi:2013ina}),
although one can make progress through EFT methods \cite{Gergely:2014rna}.
The number of free $\alpha_{i}$-like parameters is likely to increase
tremendously in addition to these models' having new unobservable
$v_{X}$-like degrees of freedom. Therefore the predictivity and the
measurability of such models is likely to be much poorer.\\

The most important conclusion of this paper is that for the gravity-like
models within our scope, there are no properties beyond those specified
above which would affect any observations. Measuring the background
$H(t)$, $\tilde{\Omega}_{\text{m0 }}$ and the four functions $\alpha_{i}$
is the maximum that can be done using linear structure formation.
If there is no evidence that the $\alpha_{i}$ are different from
zero up to some precision then dynamical dark energy does not contribute
by more than this precision to the acceleration mechanism. The question
of whether a more fundamental mechanism than $\Lambda$ is present
becomes moot, since it would be irrelevant to the dynamics. In a sense,
measuring the $\alpha_{i}$ with the best precision is the goal of
large-scale-structure measurements within the context of dark-energy
cosmology.

\section*{Acknowledgements}

The work of EB is supported by ``Fondazione Ing.~Aldo Gini'' and
``Fondazione Angelo Della Riccia''. The authors are grateful to
Luca Amendola, Guillermo Ballesteros, Bruce Bassett, Chris Clarkson,
Ruth Durrer, Bin Hu, Julien Lesgourgues, Martin Kunz, Marco Raveri,
Ippocratis Saltas, Navin Sivanandham, Alexander Vikman, Miguel Zumalacárregui
for valuable comments and criticisms. The computations
presented in this paper have been partially done with the \textit{xAct}
package for Mathematica \cite{xActwebsite,MartinGarcia2008597,Brizuela:2008ra}.

\appendix

\section{Definitions of Evolution Variables\label{sec:Definitions-of-Evolution}}

The energy density and pressure of a dark energy described by a Horndeski
Lagrangian \eqref{eq:lagrangian} are

\begin{eqnarray}
M_{*}^{2}\tilde{\mathcal{E}} & \equiv & -K+2X\left(K_{X}-G_{3\phi}\right)+6\dot{\phi}H\left(XG_{3X}-G_{4\phi}-2XG_{4\phi X}\right)\label{eq:density}\\
 &  & +12H^{2}X\left(G_{4X}+2XG_{4XX}-G_{5\phi}-XG_{5\phi X}\right)+4\dot{\phi}H^{3}X\left(G_{5X}+XG_{5XX}\right)\,,\nonumber \\
M_{*}^{2}\tilde{\mathcal{P}} & = & K-2X\left(G_{3\phi}-2G_{4\phi\phi}\right)+4\dot{\phi}H\left(G_{4\phi}-2XG_{4\phi X}+XG_{5\phi\phi}\right)\label{eq:pressure}\\
 &  & -M_{*}^{2}\alpha_{\text{B}}H\frac{\ddot{\phi}}{\dot{\phi}}-4H^{2}X^{2}G_{5\phi X}+2\dot{\phi}H^{3}XG_{5X}\,.\nonumber 
\end{eqnarray}
where we have already absorbed the contribution to the Planck mass
$M_{*}$. The equation of motion for the background values of the
scalar field \eqref{eq:BackgEoM} is an equation for the evolution
of the shift charge,
\[
\dot{n}+3Hn=\mathcal{P}_{\phi}
\]
with the charge density
\begin{eqnarray}
n & \equiv & \dot{\phi}\left(K_{X}-2G_{3\phi}\right)+6HX\left(G_{3X}-2G_{4\phi X}\right)+\label{eq:shiftcharge}\\
 &  & +6H^{2}\dot{\phi}\left(G_{4X}+2XG_{4XX}-G_{5\phi}-XG_{5\phi X}\right)+\nonumber \\
 &  & +2H^{3}X\left(3G_{5X}+2XG_{5XX}\right)\,,\nonumber 
\end{eqnarray}
and the non-conservation driven by a violation of the shift symmetry
through the term
\begin{align}
\mathcal{P}_{\phi}\equiv & K_{\phi}-2XG_{3\phi\phi}+2\ddot{\phi}\left(XG_{3\phi X}+3H\dot{\phi}G_{4\phi X}\right)+6\dot{H}G_{4\phi}+\label{eq:non-shift}\\
 & +6H^{2}\left(2G_{4\phi}+2XG_{4\phi X}-XG_{5\phi\phi}\right)+2H^{3}\dot{\phi}XG_{5\phi X\,.}\nonumber 
\end{align}
The evolution of perturbations on a particular background is determined
by four independent and dimensionless functions of time, $\alpha_{i}$,
which can be determined for any particular Lagrangian by

\begin{align}
M_{*}^{2}\equiv & 2\left(G_{4}-2XG_{4X}+XG_{5\phi}-\dot{\phi}HXG_{5X}\right)\label{eq:planckmass}\\
HM_{*}^{2}\alpha_{\textrm{M}}\equiv & \frac{\mathrm{d}}{\mathrm{d}t}M_{*}^{2}\label{eq:omega1-1}\\
H^{2}M_{*}^{2}\alpha_{\textrm{K}}\equiv & 2X\left(K_{X}+2XK_{XX}-2G_{3\phi}-2XG_{3\phi X}\right)+\label{eq:omega2}\\
 & +12\dot{\phi}XH\left(G_{3X}+XG_{3XX}-3G_{4\phi X}-2XG_{4\phi XX}\right)+\nonumber \\
 & +12XH^{2}\left(G_{4X}+8XG_{4XX}+4X^{2}G_{4XXX}\right)-\nonumber \\
 & -12XH^{2}\left(G_{5\phi}+5XG_{5\phi X}+2X^{2}G_{5\phi XX}\right)+\nonumber \\
 & +4\dot{\phi}XH^{3}\left(3G_{5X}+7XG_{5XX}+2X^{2}G_{5XXX}\right)\nonumber \\
HM_{*}^{2}\alpha_{\textrm{B}}\equiv & 2\dot{\phi}\left(XG_{3X}-G_{4\phi}-2XG_{4\phi X}\right)+\label{eq:omega3}\\
 & +8XH\left(G_{4X}+2XG_{4XX}-G_{5\phi}-XG_{5\phi X}\right)+\nonumber \\
 & +2\dot{\phi}XH^{2}\left(3G_{5X}+2XG_{5XX}\right)\nonumber \\
M_{*}^{2}\alpha_{\textrm{T}}\equiv & 2X\left(2G_{4X}-2G_{5\phi}-\left(\ddot{\phi}-\dot{\phi}H\right)G_{5X}\right)\label{eq:omega4}
\end{align}

\section{Scale Dependence in Dynamics\label{sec:Scale-Dependence-in}}

Since the value of the scalar perturbation $v_{X}$ is not an observable,
it is helpful to eliminate it by solving all the constraints in the
Einstein equations to produce an evolution equation for the Newtonian
potential $\Phi$, eq.~\eqref{eq:dynamical},
\begin{eqnarray}
\ddot{\Phi}+\frac{\beta_{1}\beta_{2}+\beta_{3}\alpha_{\textrm{B}}^{2}\frac{k^{2}}{a^{2}}}{\beta_{1}+\alpha_{\textrm{B}}^{2}\frac{k^{2}}{a^{2}}}\dot{\Phi} & + & \frac{\beta_{1}\beta_{4}+\beta_{1}\beta_{5}\frac{k^{2}}{a^{2}}+c_{S}{}^{2}\alpha_{\textrm{B}}^{2}\frac{k^{4}}{a^{4}}}{\beta_{1}+\alpha_{\textrm{B}}^{2}\frac{k^{2}}{a^{2}}}\Phi=\label{eq:dynamical-1}\\
 & = & -\frac{1}{2}\tilde{\rho}_{\text{m}}\left[\frac{\beta_{1}\beta_{6}+\beta_{7}\alpha_{\textrm{B}}^{2}\frac{k^{2}}{a^{2}}}{\beta_{1}+\alpha_{\textrm{B}}^{2}\frac{k^{2}}{a^{2}}}\delta_{\text{m}}+(1+w_{\text{m}})\frac{\beta_{1}\beta_{8}+\beta_{9}\alpha_{\textrm{B}}^{2}\frac{k^{2}}{a^{2}}}{\beta_{1}+\alpha_{\textrm{B}}^{2}\frac{k^{2}}{a^{2}}}v_{\text{m}}\right]\,,\nonumber 
\end{eqnarray}
and an anisotropy constraint that relates the two
potentials $\Phi$ and $\Psi$, eq.~\ref{eq:anisotropy},
\begin{align}
\alpha_{\textrm{B}}^{2}\frac{k^{2}}{a^{2}} & \left[\Psi-\Phi\left(1+\alpha_{\textrm{T}}+\frac{2\left(\alpha_{\textrm{M}}-\alpha_{\textrm{T}}\right)}{\alpha_{\textrm{B}}}\right)\right]+\beta_{1}\left[\Psi-\Phi\left(1+\alpha_{\textrm{T}}\right)\left(1-\frac{2DH^{2}\left(\alpha_{\textrm{M}}-\alpha_{\textrm{T}}\right)}{\beta_{1}}\right)\right]\nonumber \\
 & =\left(\alpha_{\textrm{M}}-\alpha_{\textrm{T}}\right)\left[\alpha_{\textrm{B}}\tilde{\rho}_{\textrm{m}}\delta_{\textrm{m}}-2HD\dot{\Phi}-H\left(3\alpha_{\textrm{B}}+\alpha_{\textrm{K}}\right)\left(\tilde{\rho}_{\textrm{m}}+\tilde{p}_{\textrm{m}}\right)v_{\textrm{m}}\right]\,.\label{eq:anisotropy-1}
\end{align}
Both equations are repeated here for convenience without neglecting
the matter velocity $v_{\text{m}}$. The parameters $\beta_{i}$ are
defined to be functions of the background evolution and the $\alpha_{i}$: 

\begin{eqnarray}
\beta_{1} & \equiv & \frac{3\alpha_{\textrm{B}}^{2}}{2}\left(\tilde{\rho}_{\mathrm{m}}+\tilde{p}_{\mathrm{m}}\right)-D\left[2\mathfrak{H}+\tilde{\rho}_{\mathrm{m}}+\tilde{p}_{\mathrm{m}}\right]\label{eq:beta1}\\
\beta_{2} & \equiv & 2H\left(2+\alpha_{\textrm{M}}\right)+3H\Upsilon\label{eq:beta2}\\
\beta_{3} & \equiv & H\left(3+\alpha_{\textrm{M}}\right)+\frac{\alpha_{\textrm{K}}}{D}\left(\frac{\dot{\alpha}_{\textrm{K}}}{\alpha_{\textrm{K}}}-2\frac{\dot{\alpha}_{\textrm{B}}}{\alpha_{\textrm{B}}}\right)\label{eq:beta3}\\
\beta_{4} & \equiv & \left(1+\alpha_{\textrm{T}}\right)\left[2\dot{H}+H^{2}\left(3+3\Upsilon+\alpha_{\textrm{M}}\right)\right]+\dot{\alpha}_{\textrm{T}}H\label{eq:beta4}\\
\beta_{5} & \equiv & c_{\text{s}}^{2}+\frac{\alpha_{\textrm{B}}\left(\beta_{3}-\beta_{2}\right)}{HD}+\frac{H\alpha_{\textrm{B}}^{2}}{\beta_{1}}\left(1+\alpha_{\textrm{T}}\right)\left(\beta_{3}-\beta_{2}\right)+\frac{\alpha_{\textrm{B}}^{2}\beta_{4}}{\beta_{1}}\label{eq:beta5}\\
\beta_{6} & \equiv & \beta_{7}+\frac{\alpha_{\textrm{B}}\left(\beta_{3}-\beta_{2}\right)}{HD}\label{eq:beta6}\\
\beta_{7} & \equiv & c_{\text{s}}^{2}-w_{\text{m}}+\frac{\alpha_{\textrm{B}}^{2}\left(1+\alpha_{\textrm{T}}+3w_{\text{m}}\right)+2\left(\alpha_{\textrm{M}}-\alpha_{\textrm{T}}\right)\alpha_{\textrm{B}}}{2D}\label{eq:beta7}\\
\beta_{8} & \equiv & \beta_{9}+\frac{\left(\beta_{2}-\beta_{3}\right)\left(\alpha_{\textrm{K}}+3\alpha_{\textrm{B}}\right)}{D}\label{eq:beta8}\\
\beta_{9} & \equiv & \beta_{3}-H\left(4+3c_{\text{s}}^{2}+\alpha_{\textrm{M}}+\alpha_{\textrm{T}}\right)\label{eq:beta9}
\end{eqnarray}
where we have defined
\begin{align}
\mathfrak{H} & \equiv\dot{H}-H^{2}\left(\alpha_{\textrm{M}}-\alpha_{\textrm{T}}\right)\,,\label{eq:Upsilon}\\
D & \equiv\alpha_{\textrm{K}}+\frac{3}{2}\alpha_{\textrm{B}}^{2}\,,\nonumber \\
3\beta_{1}H\Upsilon & \equiv2D\left[\dot{\mathfrak{H}}+\left(3+\alpha_{\textrm{M}}\right)H\mathfrak{H}\right]\nonumber \\
 & -H\left(\tilde{\rho}_{\textrm{m}}+\tilde{p}_{\textrm{m}}\right)\left(3\alpha_{\textrm{B}}+\alpha_{\textrm{K}}\right)\left(\alpha_{\textrm{M}}-\alpha_{\textrm{T}}\right)\nonumber \\
 & +\alpha_{\textrm{K}}\left(\dot{\tilde{p}}_{\text{m}}+\alpha_{\textrm{M}}H\tilde{p}_{\text{m}}\right)+\frac{\alpha_{\textrm{K}}\alpha_{\textrm{B}}^{2}}{2D}\left(\tilde{\rho}_{\textrm{m}}+\tilde{p}_{\textrm{m}}\right)\left(\frac{\dot{\alpha}_{\textrm{K}}}{\alpha_{\textrm{K}}}-2\frac{\dot{\alpha}_{\textrm{B}}}{\alpha_{\textrm{B}}}\right)\,,\nonumber 
\end{align}
and where the sound speed of the scalar mode is
\[
c_{\text{s}}^{2}=-\frac{\left(2-\alpha_{\textrm{B}}\right)\left[\mathfrak{H}-\frac{1}{2}H^{2}\alpha_{\textrm{B}}\left(1+\alpha_{\textrm{T}}\right)\right]-H\dot{\alpha}_{\textrm{B}}+\tilde{\rho}_{\textrm{m}}+\tilde{p}_{\textrm{m}}}{H^{2}D}\,.
\]

\section{Perturbation Equations in Synchronous Gauge\label{sec:SynchGauge}}

In this section, we provide the equations of motion for linear perturbations
in synchronous gauge and using conformal time. This is to allow for
direct implementation of our formulation in codes as CAMB \cite{Lewis:1999bs}
or CLASS \cite{Lesgourgues:2011re,Blas:2011rf}. For the metric potentials
in synchronous gauge, we continue to use the notation of ref.~\cite{Ma:1995ey}.%
\footnote{Note that CAMB redefines time derivatives of $\eta$ and $h$ using
the scheme: $h^{\prime}\rightarrow2k\mathcal{Z}$ and $\eta^{\prime}\rightarrow k/3\left(\sigma_{*}-\mathcal{Z}\right)$.%
} Note that with respect to the main text, we have made the redefintions
\begin{align*}
v_{X} & \rightarrow aV_{X}\,,\\
\mathcal{H} & \equiv aH=\frac{a'}{a}\,,
\end{align*}
for notational simplicity. The equations in this section are equivalent
to those derived in ref.~\cite{Hu:2013twa} in the effective field
theory formalism. As we discuss in section~\ref{sub:OmegasMean},
the advantage of our formulation is the fact that it is non-redundant,
splits background and perturbation contributions and is more directly
related to physical effects, as shown in section \ref{sec:Physics}.\\

\noindent The Einstein time-time equation is

\begin{align}
2k^{2}\eta & =-\frac{\rho_{\textrm{m}}\delta_{\textrm{m}}a^{2}}{{M_{*}}^{2}}+\frac{\mathcal{H}}{2}\left(2-\alpha_{\textrm{B}}\right)h^{\prime}+\mathcal{H}^{2}\left(\alpha_{\textrm{K}}+3\alpha_{\textrm{B}}\right){V_{X}}^{\prime}\label{eq:CAMB00}\\
 & +\left[\alpha_{\textrm{B}}k^{2}-3\left(\mathcal{H}^{2}-{\mathcal{H}}^{\prime}\right)\left(2-\alpha_{\textrm{B}}\right)+\mathcal{H}^{2}\left(\alpha_{\textrm{K}}+3\alpha_{\textrm{B}}\right)+3a^{2}\frac{\rho_{\textrm{m}}+p_{\textrm{m}}}{{M_{*}}^{2}}\right]\mathcal{H}V_{X}\,.\nonumber 
\end{align}

\noindent The Einstein time-space equation is 

\begin{align}
2\eta^{\prime} & =\frac{a^{2}}{{M_{*}}^{2}}\left(\rho_{\textrm{m}}+p_{\textrm{m}}\right)v_{\textrm{m}}+\alpha_{\textrm{B}}\mathcal{H}{V_{X}}^{\prime}\label{eq:CAMB0i}\\
 & -\left[2\left(\mathcal{H}^{2}-{\mathcal{H}}^{\prime}\right)-\mathcal{H}^{2}\alpha_{\textrm{B}}-a^{2}\frac{\rho_{\textrm{m}}+p_{\textrm{m}}}{{M_{*}}^{2}}\right]V_{X}\,.\nonumber 
\end{align}

\noindent The Einstein space-space traceless equation is 

\begin{align}
3\eta^{\prime\prime}+ & \frac{h^{\prime\prime}}{2}+\mathcal{H}\left(2+\alpha_{\textrm{M}}\right)\left(3\eta^{\prime}+\frac{h^{\prime}}{2}\right)-k^{2}\left(1+\alpha_{\textrm{T}}\right)\eta=\label{eq:CAMBij}\\
 & =\mathcal{H}k^{2}\left(\alpha_{\textrm{M}}-\alpha_{\textrm{T}}\right)V_{X}-\frac{a^{2}P\Pi_{\textrm{m}}}{{M_{*}}^{2}}\,.\nonumber 
\end{align}
The Einstein space-space trace equation is

\begin{align}
{h}^{\prime\prime} & =-\frac{3a^{2}}{{M_{*}}^{2}}\delta p_{\text{m}}-\mathcal{H}\left(2+\alpha_{\textrm{M}}\right)h^{\prime}+2k^{2}\left(1+\alpha_{\textrm{T}}\right)\eta-3\mathcal{H}\alpha_{\textrm{B}}{V_{X}}^{\prime\prime}+2\mathcal{H}k^{2}\left(\alpha_{\textrm{M}}-\alpha_{\textrm{T}}\right)V_{X}\label{eq:CAMBii}\\
 & -3\left[2\left({\mathcal{H}}^{\prime}-\mathcal{H}^{2}\right)+\mathcal{H}{\alpha_{\textrm{B}}}^{\prime}+\alpha_{\textrm{B}}{\mathcal{H}}^{\prime}+\mathcal{H}^{2}\alpha_{\textrm{B}}\left(3+\alpha_{\textrm{M}}\right)\right]{V_{X}}^{\prime}\nonumber \\
 & -3\left[2{\mathcal{H}}^{\prime\prime}+\mathcal{H}^{2}{\alpha_{\textrm{B}}}^{\prime}-\mathcal{H}^{3}\left(2+\alpha_{\textrm{M}}\right)\left(2-\alpha_{\textrm{B}}\right)+2\mathcal{H}{\mathcal{H}}^{\prime}\left(\alpha_{\textrm{M}}+\alpha_{\textrm{B}}\right)\right]V_{X}\nonumber \\
 & -3a^{2}\frac{\rho_{\textrm{m}}+p_{\textrm{m}}}{{M_{*}}^{2}}\left[{V_{X}}^{\prime}+\mathcal{H}V_{X}\right]-\frac{3a^{2}p_{\textrm{m}}^{\prime}}{{M_{*}}^{2}}V_{X}\,.\nonumber 
\end{align}
The scalar-field equation of motion is

\begin{align}
\mathcal{H}^{2}\left(\alpha_{\textrm{K}}+\frac{3}{2}\alpha_{\textrm{B}}^{2}\right) & {V_{X}}^{\prime\prime}+A{V_{X}}^{\prime}+\left(3B+Ck^{2}\right)V_{X}-\frac{C}{2}h^{\prime}\nonumber \\
 & +\frac{3\alpha_{\textrm{B}}}{2{M_{*}}^{2}}a^{2}\delta p_{\textrm{m}}+\left[2\left(\alpha_{\textrm{M}}-\alpha_{\textrm{T}}\right)+\alpha_{\textrm{B}}\left(1+\alpha_{\textrm{T}}\right)\right]\frac{\rho_{\textrm{m}}\mathcal{H}\delta_{\textrm{m}}a^{2}}{2{M_{*}}^{2}}=0\,,\label{eq:CAMBscalar}
\end{align}
where

\begin{eqnarray*}
A & \equiv & \frac{1}{2}\mathcal{H}^{3}\left(2-\alpha_{\textrm{B}}\right)\left[\alpha_{\textrm{K}}\left(1+\alpha_{\textrm{T}}\right)-3\alpha_{\textrm{B}}\left(1-\alpha_{\textrm{T}}+\alpha_{\textrm{M}}\right)\right]+\mathcal{H}\left(\alpha_{\textrm{K}}+\frac{3}{2}\alpha_{\textrm{B}}^{2}\right)\left(\mathcal{H}^{2}+{\mathcal{H}}^{\prime}\right)\\
 &  & +\mathcal{H}\left(\alpha_{\textrm{K}}{\mathcal{H}}^{\prime}+\mathcal{H}{\alpha_{\textrm{K}}}^{\prime}\right)+3\mathcal{H}\alpha_{\textrm{B}}\left({\mathcal{H}}^{\prime}+\frac{1}{2}\mathcal{H}{\alpha_{\textrm{B}}}^{\prime}\right)+\frac{3\left(\rho_{\textrm{m}}+p_{\textrm{m}}\right)\mathcal{H}\alpha_{\textrm{B}}a^{2}}{2{M_{*}}^{2}}\,,\\
B & \equiv & a^{2}\frac{\rho_{\textrm{m}}+p_{\textrm{m}}}{{M_{*}}^{2}}\left[{\mathcal{H}}^{\prime}+\frac{1}{2}\mathcal{H}^{2}\alpha_{\textrm{T}}\left(2-\alpha_{\textrm{B}}\right)-\mathcal{H}^{2}\left(1+\alpha_{\textrm{M}}\right)\right]+\frac{\mathcal{H}\alpha_{\textrm{B}}a^{2}p_{\textrm{m}}^{\prime}}{2{M_{*}}^{2}}\\
 &  & +\mathcal{H}^{2}\left(2-\alpha_{\textrm{B}}\right)\left[\frac{1}{6}\alpha_{\textrm{K}}\mathcal{H}^{2}\left(1+\alpha_{\textrm{T}}\right)-\mathcal{H}^{2}\left(1-\alpha_{\textrm{B}}\right)\alpha_{\textrm{T}}-{\mathcal{H}}^{\prime}\alpha_{\textrm{M}}\right]\\
 &  & +\frac{1}{2}\mathcal{H}^{2}\left(2-\alpha_{\textrm{B}}\right)^{2}\left(\mathcal{H}^{2}\alpha_{\textrm{M}}+{\mathcal{H}}^{\prime}\alpha_{\textrm{T}}\right)+\left(2-\alpha_{\textrm{B}}\right)\left(\mathcal{H}^{2}-{\mathcal{H}}^{\prime}\right)^{2}\\
 &  & +\mathcal{H}^{3}\left(\frac{1}{3}{\alpha_{\textrm{K}}}^{\prime}+{\alpha_{\textrm{B}}}^{\prime}\right)+\left(\frac{1}{2}\mathcal{H}^{2}\alpha_{\textrm{B}}-{\mathcal{H}}^{\prime}\right)\mathcal{H}{\alpha_{\textrm{B}}}^{\prime}+\left(\alpha_{\textrm{K}}+\frac{3}{2}\alpha_{\textrm{B}}^{2}\right)\mathcal{H}^{2}{\mathcal{H}}^{\prime}\,,\\
C & \equiv & \mathcal{H}{\alpha_{\textrm{B}}}^{\prime}+\left(2-\alpha_{\textrm{B}}\right)\left[\mathcal{H}^{2}\alpha_{\textrm{M}}-{\mathcal{H}}^{\prime}+\frac{1}{2}\mathcal{H}^{2}\left(2-2\alpha_{\textrm{T}}+\alpha_{\textrm{B}}+\alpha_{\textrm{B}}\alpha_{\textrm{T}}\right)\right]-a^{2}\frac{\rho_{\textrm{m}}+p_{\textrm{m}}}{{M_{*}}^{2}}\,.
\end{eqnarray*}
Note that compared to the Newtonian-gauge equation \eqref{eq:scalar},
we have diagonalised the kinetic term for the scalar by eliminating
the highest derivatives of the metric $h''$ and $\eta''$ as well
as $k^{2}\eta$ using eqs~\eqref{eq:CAMB00} and \eqref{eq:CAMBii}.
This should simplify the implementation in the Boltzmann codes. \bibliographystyle{utcaps}
\bibliography{OmegaRefs}

\end{document}